\documentclass{article}
\usepackage[a4paper,scale=0.75,centering]{geometry}

\usepackage{subfigure}

\usepackage{eufrak}
\usepackage{dsfont}
\usepackage{amssymb}%relation symbols
\usepackage{mathrsfs}

\usepackage{physics}
\newcommand{\be}{\begin{eqnarray}}
\newcommand{\ee}{\end{eqnarray}}
\usepackage{slashed} %Feynman Slash Notation

\usepackage{color}
\definecolor{ao(english)}{rgb}{0.0, 0.5, 0.0}
\usepackage[breaklinks,pdfpagelabels]{hyperref} %''hypertex'' option to compile in dvi without errors
\hypersetup{colorlinks=true,linkcolor=blue,urlcolor=blue,citecolor=ao(english)}

\usepackage{booktabs}
\usepackage{multirow}
\usepackage{threeparttable}
\usepackage{float}

\usepackage{graphicx}

\usepackage{makecell}

\usepackage[affil-it]{authblk}

\usepackage{tcolorbox}

\tcbset{colback=blue!5!white,colframe=magenta!75!black}%red/blue/green/black/white/cyan/magenta/yellow

\title{Testing black hole metrics with binary black hole inspirals}

\author[1]{Zhe~Zhao\thanks{\href{mailto:zzhao24@m.fudan.edu.cn}{zzhao24@m.fudan.edu.cn}}}
\author[1]{Swarnim~Shashank\thanks{\href{mailto:swarnim@fudan.edu.cn}{swarnim@fudan.edu.cn}}}
\author[1]{Debtroy~Das\thanks{\href{mailto:ddebtroy22@m.fudan.edu.cn}{ddebtroy22@m.fudan.edu.cn}}}
\author[1,2]{Cosimo~Bambi\thanks{\href{mailto:bambi@fudan.edu.cn}{bambi@fudan.edu.cn} (Corresponding author)}}

\affil[1]{Center for Astronomy and Astrophysics, Center for Field Theory and Particle Physics, and Department of Physics, Fudan University, Shanghai 200438, China}
\affil[2]{School of Natural Sciences and Humanities, New Uzbekistan University, Tashkent 100007, Uzbekistan}

\date{\today}%\today

\begin{document}
\maketitle 
\begin{abstract} 
Gravitational wave astronomy has opened an unprecedented window onto tests of gravity and fundamental physics in the strong-field regime. In this study, we examine a series of well-motivated deviations from the classical Kerr solution of General Relativity and employ gravitational wave data to place constraints on possible deviations from the Kerr geometry. 
The method involves calculating the phase of gravitational waves using the effective one-body formalism and then applying the parameterized post-Einsteinian framework to constrain the parameters appearing in these scenarios beyond General Relativity. The effective one-body method, known for its capability to model complex gravitational waveforms, is used to compute the wave phase, and the post-Einsteinian framework allows for a flexible, model-independent approach to parameter estimation. 
We demonstrate that gravitational wave data provide evidence supporting the Kerr nature of black holes, showing no significant deviations from General Relativity, thereby affirming its validity within the current observational limits. 
We further assess the potential impact of orbital eccentricity and find that, within observationally allowed ranges, its contribution to the inferred deviations is subdominant. This work bridges theoretical waveform modeling with observational constraints, providing a pathway to test the no-hair theorem and probe the astrophysical viability of modified black holes.

\end{abstract}

\section{Introduction}

% GW is a way for detecting new physics
Gravitational waves (GWs) are ripples in spacetime caused by massive accelerating objects such as black holes (BHs)~\cite{Wald:1984rg,Liang:2023ahd}. Since their first detection by the LIGO and Virgo Collaboration in 2015~\cite{LIGOScientific:2016aoc}, GWs have revolutionized our understanding of astrophysical phenomena, particularly those involving BHs and neutron stars~\cite{LIGOScientific:2017vwq}. In General Relativity (GR) and in the absence of exotic matter fields, the spacetime metric around astrophysical BHs is expected to be approximated well by the Kerr solution, which is completely specified by two parameters: the BH mass and the BH spin angular momentum. The detection of GWs from binary BH mergers offers a unique opportunity to probe the properties of these elusive objects, including possible deviations from the Kerr solution~\cite{Maggiore:2007ulw}.

%EOB: computation of phase  
To constrain the deviations from the Kerr metric, we adopt using GW phase information, which has been shown to provide more accurate measurements than amplitude data~\cite{Perkins:2022fhr}. The effective one-body (EOB) formalism is an ideal framework for modeling the complex dynamics of BH mergers, as it can compute the phase evolution of the system~\cite{Damour:2008yg, Blanchet:2013haa}. The EOB method, originally formulated without considering spin~\cite{Buonanno:1998gg}, was extended to incorporate BH spin in Ref.~\cite{Damour:2001tu}, a crucial aspect for understanding the behavior of real astrophysical systems. Studies have demonstrated that including spin effects may be important for accurately modeling gravitational waveforms and extracting reliable parameter estimates~\cite{Chandramouli:2024vhw}.

%ppE: parameterization of phase of GW
The parameterized post-Einsteinian (ppE) framework, a versatile approach for testing GR and theories beyond GR, provides a way to place constraints on possible deviations from the Kerr geometry. The ppE framework is particularly valuable because it allows us to test deviations from GR without making strong assumptions about the underlying physics~\cite{Yunes:2009ke}.

% why rotating metric 
Previous studies have used similar methods to extract parameters for gravitational waveforms from other theoretical models, without including BH spin effects~\cite{Cardenas-Avendano:2019zxd,Shashank:2021giy} and including BH spin effects~\cite{Das:2024mjq}. 
We use the same method for constraining possible deviations from the Kerr geometry as in Refs.~\cite{Cardenas-Avendano:2019zxd,Shashank:2021giy,Das:2024mjq} by utilizing the EOB formalism for phase calculations and the ppE framework for parameter extraction. By combining these two powerful techniques, we aim to provide stringent bounds on a number of ``deformation parameters'', contributing to further understanding of the fundamental properties of BHs. These deformation parameters are parameters that appear in these BH spacetimes beyond GR and can be used to measure possible deviations from the Kerr solution. As GW detectors continue to improve, this approach will play a critical role in testing GR and exploring the potential for new physics beyond the Einsteinian framework~\cite{LISA:2022yao}.

%essence of work
It is important to note that our study does not, in essence, provide anything new. The analysis of GWs has already been carried out by the LIGO-Virgo Collaboration and no deviations from the predictions of GR have been detected. Our work maps their results to specific BH metrics, similar to the studies reported in~\cite{Cardenas-Avendano:2019zxd,Shashank:2021giy}, thereby obtaining constraints on a number of BH spacetimes beyond GR. In addition, we assess the potential impact of unmodeled physical effects, such as orbital eccentricity, on the inferred deformation parameters. This allows us to interpret the observational results in a physically transparent framework and to evaluate the robustness of the resulting constraints.

% structure of paper 
The paper is structured as follows. In Sec.~\ref{sec:methodology}, we discuss the methodology and assumptions involved in the computations of the phase corrections due to the deviations from the Kerr spacetime. Sec.~\ref{sec:results} is organized into numerous subsections, each dedicated to a specific model and fundamental physics scenario examined. In these subsections, we constrain BH spacetimes beyond GR by comparing the theoretically predicted GW phases with observational data from the GW170608 event. %In Sec.~\ref{sec:briefdiscussion}, we present a concise summary of our results.
In Sec.~\ref{sec:briefdiscussion}, we summarize our main results and discuss potential systematic effects, in particular the impact of orbital eccentricity on the estimation of deformation parameters. Our conclusions are reported in Sec.~\ref{sec:conclusions}.

\section{Methodology}
\label{sec:methodology}

% in general, conservative and dissipative sectors ----> phase 
To compute the GW phase during the inspiral phase of binary BHs in a general modified gravity theory, we note that it is, in principle, fully determined by modifications to both the conservative and dissipative sectors of the binary system~\cite{Yunes:2009ke,Blanchet:2006zz}.

For the modification to the conservative sector within the modified gravity theory, we postulate its computation via the EOB approach. Specifically, to account for the spin effects of BHs, we adopt the EOB formalism that incorporates spin. We assume that the conservative dynamics of spinning binary BHs during the inspiral phase are governed by a $\nu$-deformed Kerr metric~\cite{Damour:2001tu}. However, in our calculations, similar to \cite{Cardenas-Avendano:2019zxd}, we will use the BH spacetimes from models beyond GR, rather than the $\nu$-deformed metrics.

%dissipative dynamics remains described by GR
Determining the GW phasing additionally requires knowledge of the modifications to the dissipative sector. Given that the leading-order contribution in the modified gravity theory is GR, and considering that the order at which the modified gravity corrections enter generally differs from the order governing the dissipative effects, we assume that the order associated with the modified gravity corrections is not less than that governing the dissipative sector~\cite{Cardenas-Avendano:2019zxd}. Consequently, within this assumption, the dissipative dynamics remains described by GR, i.e., the energy loss rate satisfies $\dot E\simeq -\mathcal{L}_{\rm GW}$.
Should the dissipative corrections enter at a higher post-Newtonian (PN) order than the modified gravity terms, our calculations would yield more conservative constraints on the parameters of the modified theory.

\subsection{Special circular metrics}

The metrics considered in this paper are special circular metrics, which can be written in the Boyer-Lindquist coordinate system $\{t, r, \theta, \varphi\}$ as:
\be\label{circular-metric} g_{ab}=g_{tt}\qty(\dd t)_a\qty(\dd t)_b+g_{t\varphi}\qty(\dd t)_a\qty(\dd \varphi)_b+g_{\varphi t}\qty(\dd \varphi)_a\qty(\dd t)_b+g_{rr}\qty(\dd r)_a\qty(\dd r)_b+g_{\theta\theta}\qty(\dd \theta)_a\qty(\dd \theta)_b+g_{\varphi\varphi}\qty(\dd \varphi)_a\qty(\dd \varphi)_b \,,\ee
where
\be \pdv{g_{\mu\nu}}{t}=\pdv{g_{\mu\nu}}{\varphi}=0\,,\ee
i.e., \be  \qty(\pdv{t})^a\qand\qty(\pdv{\varphi})^a\ee are Killing vector fields of the circular spacetime we are considering \eqref{circular-metric}. 

We consider the theories (or metrics) listed in Sec.~\ref{sec:results}, whose metrics have the form similar to \eqref{circular-metric}. We examine the case where they deviate slightly from the Kerr metric and use the deformation parameters $\{\delta_i\}$ to characterize these deviations.

\subsection{Effective one body problem}
\label{sec.method}

In the PN formalism, the two-body problem is reformulated as an EOB problem, where a test particle with reduced mass $\mu = m_1 m_2 / m$ (where $m = m_1 + m_2$, and $m_1$ and $m_2$ denote the masses of the two bodies) moves along a geodesic trajectory around an object of mass $m$. The four-momentum of the test particle is contracted with the metric \eqref{circular-metric}, leading to the effective Hamiltonian that governs the conservative dynamics of the orbital motion~\cite{Buonanno:1998gg, Hinderer:2017jcs}. This effective model can subsequently be mapped back to the original two-body problem, yielding the GWs emitted by the parametrically deformed binary BH system.

\subsubsection{Equatorial circular geodesics of massive particles}
\label{sec:c-orbit-EL}

The four-velocity of the geodesic particle moving in the equatorial plane\footnote{If the metric is invariant under the transformation \(\theta \to \pi - \theta\), then the motion of the particle with \(\eval{\theta}_i = \pi/2\) and \(\eval{\dv*{\theta}{\tau}}_i = 0\) will be confined to the equatorial plane.} is
\be\label{four-velocity} \qty(\pdv{\tau})^a= \dv{t}{\tau}\qty(\pdv{t})^a+\dv{r}{\tau}\qty(\pdv{r})^a+\dv{\varphi}{\tau}\qty(\pdv{\varphi})^a\,,\ee
and the normalization condition for tangent vectors is equivalent to:
\be\label{normalization condition} -1=g_{ab}\qty(\pdv{\tau})^a\qty(\pdv{\tau})^b=\qty(\dv{t}{\tau})^2g_{tt}+\qty(\dv{\varphi}{\tau})^2g_{\varphi\varphi}+\qty(\dv{r}{\tau})^2g_{rr}+2\dv{t}{\tau}\dv{\varphi}{\tau}g_{t\varphi}\,.\ee

Additionally, we can define two conserved quantities for the massive particle, namely the ``energy'' and the ``angular momentum'' as:\be \label{E amd L} E=-g_{ab}\qty(\pdv{t})^a\qty(\pdv{\tau})^b=-\dv{t}{\tau}g_{tt}-\dv{\varphi}{\tau}g_{t\varphi}\qand L=g_{ab}\qty(\pdv{\varphi})^a\qty(\pdv{\tau})^b=\dv{\varphi}{\tau}g_{\varphi\varphi}+\dv{t}{\tau}g_{\varphi t}\,.\ee 
We can solve for $\dv*{\varphi}{\tau}$ and $\dv*{t}{\tau}$ from Eq.~\eqref{E amd L}:\be \label{inver-solve}\dv{t}{\tau}=-\frac{Eg_{\varphi\varphi}+Lg_{t\varphi}}{g_{tt}g_{\varphi\varphi}-g_{t\varphi}^2}\qand \dv{\varphi}{\tau}=\frac{Eg_{\varphi t}+Lg_{tt}}{g_{tt}g_{\varphi\varphi}-g_{t\varphi}^2}\,.\ee

By substituting \eqref{inver-solve} into the normalization condition \eqref{normalization condition}, we can obtain the equation for radial motion: \be -1=L \dv{\varphi}{\tau}-E\dv{t}{\tau}+\qty(\dv{r}{\tau})^2g_{rr}=g_{rr}\qty(\dv{r}{\tau})^2+\frac{E^2g_{\varphi\varphi}+L^2g_{tt}+2ELg_{t\varphi}}{g_{tt}g_{\varphi\varphi}-g_{t\varphi}^2}\,.\ee This equation can be simplified to:
\be\label{V-eff} \qty(\dv{r}{\tau})^2+V_{\text{eff}}=0\qand V_{\text{eff}}=\frac{E^2g_{\varphi\varphi}+L^2g_{tt}+2ELg_{t\varphi}}{g_{rr}\qty(g_{tt}g_{\varphi\varphi}-g_{t\varphi}^2)}+\frac{1}{g_{rr}}\,.\ee 
The equatorial circular geodesics condition can be written as: \be\label{circular-orbit} \dot r=0\Longleftrightarrow V_{\text{eff}}=0\qand \dv{r}V_{\text{eff}}=0 \,.\ee

\subsubsection{GW phase corrections induced by deviations from the Kerr geometry}

For a massive test particle in an equatorial circular orbit, the effective potential \( V_{\text{eff}} \) is a function of the radius $r$, the particle's energy $E$, the particle's angular momentum $L$, and the parameters of the BH spacetime (mass, spin, and possible deformation parameters). By solving Eq.~\eqref{circular-orbit}, we can obtain the energy $E$ and angular momentum $L$ corresponding to the circular orbit in the equatorial plane.

%\omega 

In the far-field limit, the angular velocity of the test particle, $\omega$, can be written as: 
\be\label{eq.omega-r} \omega^2=\frac{L^2}{r^4}\simeq f(r,a,\delta_i) \,,\ee
where $a$ is the black hole spin parameter and $f(r,a,\delta_i)$ contains only the zero-order and first-order corrections to the angular velocity due to the deformation parameter(s).
%E(\omega)
By expressing $r$ in Eq.~\eqref{eq.omega-r} as a function of $\omega$ and the deformation parameter(s), we can rewrite the energy $E$ of the unit-mass test particle as a function of $\omega$ and the deformation parameter(s). We introduce $\mathcal{E}=\mu E$, where $\mu$ is still the reduced mass of the system.   
%Orbital phase 
The orbital frequency is  $\nu=\omega/2\pi$ and the orbital phase is given by
\be\label{def-phase} \phi(\nu)=\int^\nu\omega\dd t=\int^\nu\frac{1}{\mathcal{\dot E}}\qty(\dv{\mathcal{E}}{\omega})\omega\dd\omega\,.\ee
Since we are only concerned with corrections to the conservative sector and assume that the dissipative sector remains unmodified compared to GR, in order to calculate the leading-order deviation in GR, as done in \cite{Cardenas-Avendano:2019zxd}, we only need to use the quadrupole formula to model the change in the binding energy:
\be \mathcal{\dot E}\simeq \mathcal{\dot E}^{0\rm PN}_{\rm GR}=-\frac{32}{5}\eta^2m^2r^4\omega^6\,,\ee where $\eta = \mu/m$ is symmetric mass ratio.

We write the orbital phase as
\be \phi(\nu)=\phi_{\rm GR}(\nu)+ \phi_{\rm NGR}(\nu)+\order{\delta_i{}^2} \,,\ee
where $\phi_{\rm GR}(\nu)$ is the GR orbital phase and $\phi_{\rm NGR}(\nu)$ contains only the  leading order correction to the phase due to the deformation parameter(s).
The Fourier transform of $\phi$ in the stationary phase approximation is $\Psi_{\rm GW}=2\phi(t_0)-2\pi f t_0$, where $t_0$ is the stationary time such that $\nu(t_0)=f/2$ and $f$ is the Fourier frequency~\cite{Cardenas-Avendano:2019zxd}. 
At the leading order in the deformation parameter(s), we have
\be\label{eq.phase.correction.general} \Psi_{\rm GW}(u)=\Psi_{\rm GW}^{\rm GR}(u) +\Psi_{\rm GW}^{\rm NGR}(u) \,,\ee
where $u=\eta^{3/5}\pi mf$ and $\Psi_{\rm GW}^{\rm NGR}(u)$ contains only the leading order correction to the phase due to the deformation parameter(s).

\subsection{Constraining the deformation parameter from binary BH events}
In this paper, we consider only BH spacetimes with one deformation parameter. For convenience in notation, from now on we will use $\delta$ to represent any of the elements in the set $\{\delta_i\}$.

%ppE
To obtain constraints on the deformation parameters from GW data, we need to compare the leading order phase correction \eqref{eq.phase.correction.general} to the ppE  framework~\cite{Khan:2015jqa}, where
\be\label{eq.ger-term-ngr-phase} \Psi_{\rm GW}^{\rm NGR}=\beta u^b \,,\ee
at the $i/2$ PN order \be\label{eq.ger-exp-beta} \beta=\frac{3}{128} \varphi_i\delta\varphi_i \eta^{-\frac{i}{5}} \,,\ee and \be\label{eq.ger-b-i} b={\frac{i-5}{3}}\,.\ee 
$\varphi_i$ is the PN phase and has the form \be\label{eq.i=0} \varphi_0&=&1\\\label{eq.i=1} \varphi_1&=&0\\\label{eq.i=2} \varphi_2 &=& \frac{3715}{756}+\frac{55 \eta }{9} \\\label{eq.i=3} \varphi_3 &=& -16 \pi +\frac{113 \delta  \chi
_a}{3}+\qty(\frac{113}{3}-\frac{76 \eta }{3}) \chi _s \\\label{eq.i=4}\varphi_4 &=& \frac{15293365}{508032}+\frac{27145 \eta }{504}+\frac{3085 \eta^2}{72}+\qty(-\frac{405}{8}+200 \eta)\chi _a^2-\frac{405}{4} \xi \chi _a \chi _s+\qty(-\frac{405}{8}+\frac{5 \eta }{2}) \chi _s^2 
\,. \ee
The individual masses and spin parameters, $m_i$ and $\chi_i$ ($i=1,2$), are encoded in the following parameter combinations \be  
m = m_1 + m_2, \\   \eta = m_1m_2 /m^2, \\   
\xi = (m_1 - m_2) /m, \\ \chi_s = (\chi_1 + \chi_2)/2, \\   \chi_a = (\chi_1 - \chi_2)/2 \,.\ee

By comparing \eqref{eq.phase.correction.general} and \eqref{eq.ger-term-ngr-phase}, it turns out that we can express the deformation parameter in the following form: \be\label{eq.def.delta} \delta= K \varphi_i\delta\varphi_i\ee where $K$ is a constant.

Under the approximation detailed in this section, it becomes feasible to compute the phasing corrections during the inspiral phase for generic modified gravity theories. In the next section, we will constrain the deformation parameters of various gravitational models using data from GW170608 with SEOBNRv4P waveform model at the 90\% confidence level. We consider GW170608 because our method is sensitive only to the inspiral phase and GW170608, with the low values of the masses of its two BHs, has a long detected inspiral phase and provides the most stringent constraints on possible deviations from the Kerr metric.

\section{Results}
\label{sec:results}

We consider the following BH spacetimes, whose metrics have the form similar to \eqref{circular-metric}. We examine the case where they deviate slightly from the Kerr metric and use the deformation parameter to characterize this deviation.

The scenarios we consider, along with the fundamental parameters we constrain, include various regular BHs, novel frameworks of fundamental physics, and BH analogs involving well-motivated structures. %such as wormholes and naked singularities.
\begin{itemize}
	\item Kerr-Newman BH \S.\ref{sec.KN}
	\item Bardeen magnetically charged regular BH \S.\ref{sec.Bardeen}
	\item Asymptotically safe gravity BH \S.\ref{sec.Asymptoticallysafe}
	\item Loop quantum gravity BH \S.\ref{sec.lqg}
	\item Non-commutative geometry BH \S.\ref{sec.NCG}
	\item Kerr-Sen BH \S.\ref{sec.KS}
	\item Ghosh-Kumar BH \S.\ref{sec.Ghosh-Kumar}
	\item Rotating Kazakov-Solodukhin BH \S.\ref{sec.Rotating Kazakov-Solodukhin}
	\item Ghosh regular BH \S.\ref{sec.Ghosh}
	\item Tinchev BH \S.\ref{sec.Tinchev}
	\item Rotating Einstein-Born-Infeld BH \S.\ref{Rotating EBI}
	\item Balart-Vagenas BHs \S.\ref{sec.other}
\end{itemize}

At the end of each subsection, we not only present our constraints on the deformation parameters but also list the results (if available) that constrain these metrics through BH imaging and X-ray observations.

\subsection{Kerr-Newman black hole}
\label{sec.KN}

%whice metric and deformation parameter
The first geometry we consider beyond the Kerr solution, and arguably the physical thereof, is the Kerr-Newman (KN) spacetime. We treat the charge parameter of the KN BH as a deformation parameter and calculate the resulting modification to the binary inspiral phasing imposed by this parameter\footnote{Here we consider the charge parameter of the KN metric as a ``dark'' charge from a dark sector~\cite{Gupta:2021rod}, so there are no electromagnetic effects associated to this charge.}.

%line element
The KN metric in Boyer-Lindquist coordinates \{$t,r,\theta,\varphi$\} can be written as~\cite{Kerr:1963ud,Newman:1965tw}
\be\label{eq.le.KN}
\dd s^2  &=& - \frac{\Delta -a^2\sin^2\theta}{\rho^2} \dd t^2+ \frac{\rho^2}{\Delta} \dd r^2+ \rho^2 \dd\theta^2\nonumber \\ &&  +\frac{\qty(r^2+a^2)^2-\Delta a^2\sin^2\theta}{\rho^2}\sin^2\theta \dd\varphi^2-\frac{2a\qty(r^2+a^2-\Delta) \sin^2\theta}{\rho^2} \dd t \dd\varphi\, ,
\ee
where
\be \label{kn-rho}
\rho^2 = r^2 + a^2 \cos^2\theta\qand
\Delta = r^2 - 2mr + a^2 + q^2m^2\,.
\ee
The cosmic censorship hypothesis requires that spacetime does not contain naked singularities, which is equivalent to the condition \be\label{eq.kn.delta.condition} a^2+q^2m^2< m^2  \,.\ee

Following the approach in Sec.~\S.\ref{sec:methodology} \cite{Cardenas-Avendano:2019zxd}, we first compute the energy and angular momentum of a test particle in the KN spacetime \eqref{eq.le.KN}. Subsequently, the energy is expressed as a function of the orbital angular frequency. Finally, utilizing the phasing formula within the stationary phase approximation, the modification to the binary inspiral phasing induced by the charge parameter $q$ can be expressed as
\be\label{eq.kn.phase.correction} \Psi_{\rm GW}^{\rm NGR}(f)=-\frac{5 }{48 u \eta^{2/5}}q^2 \,,\ee
where $u=\eta^{3/5}\pi mf$.
Utilizing the ppE parameterization given in Eq.~\eqref{eq.ger-term-ngr-phase}, this phase correction \eqref{eq.kn.phase.correction} can be expressed as\be \label{eq.kn.beta.exp}\beta=-\frac{5 }{48\eta^{2/5}}q^2 \,.\ee
The modification enters at $b=-1$, corresponding to the $i=2$ and $1$PN order. 
Therefore, to facilitate comparison with GW observations, we can equivalently combine Eqs.~\eqref{eq.ger-exp-beta} and \eqref{eq.kn.beta.exp} to find that 
\be\label{eq.kn.phase.1} q^2=-\frac{9}{40}\varphi_2\delta\varphi_2\,.\ee

%% constrain the deformation parameters
When using GW170608 data with SEOBNRv4P waveform model to constrain the charge parameter, we should take into account the natural constraint on the charge parameters given by \eqref{eq.kn.delta.condition}. For convenience, we use $0<q^2 < 1$, which leads to more conservative constraint. %analysis
By comparing this theoretical prediction with gravitational wave data, we constrain the dimensionless charge parameter to  $q^2=0.13^{+0.35}_{-0.13}$ at the 90\% confidence level.

%other constraint method of deformation parameters 
Constraints on $q$ from BH imaging were reported in Ref.~\cite{Vagnozzi:2022moj}. For non-rotating charged BHs, specifically Reissner-Nordstr\"om (RN) BHs, observations from the Event Horizon Telescope (EHT) impose a $1\sigma$ upper limit on the charge, $q < 0.8$, and a $2\sigma$ upper limit, $q < 0.95$. Consequently, with more than $2\sigma$ significance, the EHT data exclude the possibility of Sagittarius $\rm A^*$ being an extremal RN BH ($q = 1$). Furthermore, the naked singularity regime is unequivocally ruled out, eliminating the scenario in which Sagittarius $\rm A^*$ could represent one of the simplest forms of naked singularities~\cite{Bambi:2008jg}.

Building upon the constraints obtained from GWs and BH shadow observations, we now turn to an additional probe: X-ray reflection spectroscopy~\cite{Bambi:2016sac,Abdikamalov:2019yrr,Abdikamalov:2020oci,Bambi:2015kza}. This technique, which analyzes the reflection features in the X-ray spectrum emitted by the accretion disk of a BH, provides an alternative method for constraining BH parameters, including those related to spacetime deformations. By comparing the observed spectra to theoretical models, it is possible to place limits on the deformation parameters, such as those indicative of deviations from the Kerr metric~\cite{Cao:2017kdq,Tripathi:2018lhx,Tripathi:2019bya,Tripathi:2020yts}.
To the best of our knowledge, this method has not yet been employed to constrain the charge parameter of the KN spacetime.

\subsection{Bardeen black hole}
\label{sec.Bardeen}

%motivation and the deformation parameter

We now focus on regular BH solutions, which avoid the singularity at $r = 0$ present in the KN metric. 
The Bardeen spacetime~\cite{Bardeen(1968)}, one of the earliest such solutions, describes a regular BH with a de Sitter core. 
Carrying a magnetic charge $q$, the Bardeen BH can be seen as a magnetic monopole satisfying the weak energy condition~\cite{Ayon-Beato:2000mjt}. It also arises in GR, through a specific nonlinear electrodynamics Lagrangian~\cite{Ayon-Beato:2004ywd}, and can further be interpreted as a Schwarzschild BH with quantum corrections~\cite{Maluf:2018ksj}.
The metric of non-rotating Bardeen magnetically charged regular BH can be written  as~\cite{Bardeen(1968)} \be g_{tt}=-\qty[1-\frac{2mr^2}{\qty(r^2+q^2m^2)^{3/2}}]  \,,g_{rr}= \qty(-g_{tt})^{-1} \,,g_{\theta\theta}=r^2\,,g_{\varphi\varphi}=r^2\sin^2\theta \,.\ee The regularity requires that the spacetime does not contain singularities, which is equivalent to the condition~\cite{Bardeen(1968)} \be\label{eq.bardeen.delta.condition}  q\leq \sqrt{16/27}\approx 0.77\,.\ee
Since the Bardeen BH reduces to the Schwarzschild BH when $q = 0$, we choose $q$ as the deformation parameter and calculate the resulting modification to the binary inspiral phasing imposed by this parameter

Using the Newman-Janis transformation, we can obtain the rotating Bardeen BH in Boyer-Lindquist coordinates \{$t,r,\theta,\varphi$\}, which can be written as~\cite{Bambi:2013ufa}
\be\label{eq.le.br} \dd s^2  &=& - \frac{\Delta -a^2\sin^2\theta}{\rho^2} \dd t^2+ \frac{\rho^2}{\Delta} \dd r^2+ \rho^2 \dd\theta^2\nonumber \\ &&  +\frac{\qty(r^2+a^2)^2-\Delta a^2\sin^2\theta}{\rho^2}\sin^2\theta \dd\varphi^2-\frac{2a\qty(r^2+a^2-\Delta) \sin^2\theta}{\rho^2} \dd t \dd\varphi\, , \ee 
where
\be  \rho^2 = r^2 + a^2 \cos^2\theta\qand \Delta = r^2 - 2m(r)r + a^2\,, \ee and 
\be m(r)=m\frac{r^3}{\qty(r^2+q^2m^2)^{3/2}}\,.\ee

%{\color{red} phase correction }

We proceed as in the case of the KN spacetime. First, we compute the energy and angular momentum of a test particle in the rotating Bardeen BH \eqref{eq.le.br}. The energy is then expressed as a function of the orbital angular frequency. Finally, utilizing the phasing formula within the stationary phase approximation, the modification to the binary inspiral phasing induced by the deformation parameter $q$ can be expressed as
\be\label{eq.Bardeen.phase.correction} \Psi_{\rm GW}^{\rm NGR}(f)= -\frac{225 }{64 u^{1/3} \eta^{4/5}}q^2  \,,\ee
where $ u=\eta^{3/5}\pi mf$.
Utilizing the ppE parameterization given in Eq.~\eqref{eq.ger-term-ngr-phase}, this phase correction \eqref{eq.Bardeen.phase.correction} can be expressed as\be \label{eq.Bardeen.beta.exp}\beta=- \frac{225 }{64 \eta^{4/5}}q^2 \,.\ee
The modification enters at $b=-1/3$, corresponding to the $i=4$ and $2$PN order. 
Therefore, to facilitate comparison with GW observations, we can equivalently combine Eqs.~\eqref{eq.ger-exp-beta} and \eqref{eq.Bardeen.beta.exp} to find that 
\be\label{eq.Bardeen.phase.1} q^2=-\frac{1}{150}\varphi_4\delta\varphi_4\,.\ee

When using GW170608 data with SEOBNRv4P waveform model  to constrain the deformation parameter, we should take into account the physical restriction on the deformation parameters given by \eqref{eq.bardeen.delta.condition}. The charge constraint at the 90\% confidence level is $q^2=0.12^{+0.40}_{-0.12}$, implying that $q^2<0.52$. This result is consistent with GR within the 90\% confidence interval, showing no significant deviation.

Horizon-scale BH images enable testing gravity in strong-field regimes. We constrain deviations from GR using  EHT observations of Sagittarius $\rm A^*$, linking the bright ring size to the underlying BH shadow and precise mass-to-distance measurements.
Using this method, there are no constraints from current data on the $q$ parameter of the non-rotating Bardeen BH at both $1\sigma$ and $2\sigma$ confidence levels~\cite{Vagnozzi:2022moj}.

We are not aware of any attempt to constrain the Bardeen charge $q$ with X-ray reflection spectroscopy. However, the study reported in Ref.~\cite{Bambi:2014nta} shows that we can infer the $3\sigma$ bound $q < 0.41$ from the thermal spectrum of the accretion disk of Cygnus~X-1.

\subsection{Asymptotically safe gravity}
\label{sec.Asymptoticallysafe}

The renormalization problem in gravitational theory represents a significant obstacle in the development of a quantum theory of gravity. Asymptotically safe gravity offers a quantum realization of scale symmetry, presenting a promising avenue for the ultraviolet extension of GR. The metric of the first renormalization group improved BH spacetime, within the framework of asymptotically safe gravity, can be expressed as~\cite{Bonanno:2000ep}:
\be -g_{tt}=(g_{rr})^{-1}=1-\frac{2m}{r}G(r) \,,g_{\theta\theta}=r^2\,,g_{\varphi\varphi}=r^2\sin^2\theta \,,\ee where \be G(r)=\frac{r^3}{r^3+\tilde\omega m^2(r+\gamma m)}\,.\ee If \( \tilde{\omega} = 0 \) and \( \gamma \neq 0 \), this metric will degenerate into the Schwarzschild BH. Here we consider the case where \( \tilde{\omega} \neq 0 \) and \( \gamma = 0 \).
We treat the $\tilde{\omega}$ of the asymptotically safe BH as the deformation parameter and calculate the resulting modification to the binary inspiral phasing imposed by this parameter. The deformation parameter $\tilde{\omega}$ is in principle a free parameter, but as discussed in the literature \cite{Vagnozzi:2022moj}, we focus on the region
\be\label{eq.asg.delta.condition} 0<\tilde \omega<1  \,,\ee
where BHs have two event horizons.

Using the Newman-Janis transformation, we can obtain the rotating  asymptotically safe BH metric in Boyer-Lindquist coordinates \{$t,r,\theta,\varphi$\}, which can be written as~\cite{Torres:2017gix}
\be\label{eq.le.asg} \dd s^2  &=& - \frac{\Delta -a^2\sin^2\theta}{\rho^2} \dd t^2+ \frac{\rho^2}{\Delta} \dd r^2+ \rho^2 \dd\theta^2\nonumber \\ &&  +\frac{\qty(r^2+a^2)^2-\Delta a^2\sin^2\theta}{\rho^2}\sin^2\theta \dd\varphi^2-\frac{2a\qty(r^2+a^2-\Delta) \sin^2\theta}{\rho^2} \dd t \dd\varphi\, , \ee 
where
\be  \rho^2 = r^2 + a^2 \cos^2\theta\qand \Delta = r^2 - 2m(r)r + a^2\,, \ee and 
\be m(r)=m\frac{r^2}{r^2+\tilde\omega m^2}\,.\ee

The modification to the binary inspiral phasing induced by the deformation parameter $\tilde\omega$ can be expressed as
\be\label{eq.asg.phase.correction} \Psi_{\rm GW}^{\rm NGR}(f)= - \frac{75}{32 u^{1/3} \eta^{4/5}}\tilde\omega \,,\ee
where $u=\eta^{3/5}\pi mf$.
Utilizing the ppE parameterization, this phase correction \eqref{eq.asg.phase.correction} can be expressed as\be \label{eq.asg.beta.exp}\beta=- \frac{75}{32 \eta^{4/5}}\tilde\omega \,.\ee
The modification enters at $b=-1/3$, corresponding to the $i=4$ and $2$PN order. 
From Eqs.~ \eqref{eq.ger-exp-beta} and \eqref{eq.asg.beta.exp}, we find
\be\label{eq.asg.phase.1}\tilde \omega=-\frac{1}{100} \varphi_4\delta\varphi_4\,.\ee

When using GW170608 data with SEOBNRv4P waveform model  to constrain the deformation parameter $\tilde\omega$, we should take into account the physical restriction on the $\tilde\omega$ given by \eqref{eq.asg.delta.condition}. The constraint at the 90\% confidence level is $\tilde\omega =0.18^{+0.60}_{-0.18}$, implying that $\tilde\omega<0.78$. This result is consistent with GR within the 90\% confidence interval, showing no significant deviation.

From the EHT image of Sagittarius $\rm A^*$, we can constrain the deformation parameter of the nonrotating asymptotically safe BH: $\tilde{\omega} < 0.9$ ($1\sigma$)~\cite{Vagnozzi:2022moj}. From X-ray reflection spectroscopy, the analysis of an observation of the BH X-ray binary GRS~1915+105 provides the tighter constraints $\tilde{\omega} < 0.05$ at the 90\% of confidence level~\cite{Zhou:2020eth}.

\subsection{Loop quantum gravity}

\label{sec.lqg}

Loop quantum gravity (LQG) is a leading approach to quantum gravity, aiming to construct a quantum Riemannian geometry beyond classical spacetimes. It proposes a granular structure of spacetime, characterized by a minimum length on the order of the Planck length. This structure is composed of Planck-scale loops, whose network and dynamics are described by spin networks and spin foams, respectively. By solving the effective equations of LQG, the quantum extension of the Schwarzschild metric was derived in previous works~\cite{Bodendorfer:2019nvy,Bodendorfer:2019jay}.

Using the Newman-Janis transformation, we can obtain the rorating loop quantum gravity BH in Boyer-Lindquist coordinates \{$t,r,\theta,\varphi$\}, which can be written as~\cite{Brahma:2020eos}
\be \label{eq.le.lqg} \dd s^2 = -  \qty( 1-\frac{2 M b}{\rho^2} ) \dd t^2	-\frac{4aMb\sin^2{\theta}}{\rho^2}\dd t \dd\varphi +\rho^2\dd\theta^2+\frac{\rho^2 \dd r^2}{\Delta}	+\frac{\varSigma \sin^2{\theta}}{\rho^2}\dd \varphi^2\,,\ee where  \be &&\rho^2 = b^2 + a^2 \cos^2{\theta}\qand  	M = b\qty(1 - 8A_\lambda M^{2}_{b} \tilde{a})/2 , \\ &&\Delta = 8A_\lambda M^{2}_{b} \tilde{a} b^2 + a^2 \qand	\Sigma = (b^2 + a^2)^2 - a^2 \Delta \sin^2{\theta}, \ee and the terms $ \tilde{a} $ and $ b $ are defined as \be  b(r)^2 &=& \frac{512A_{\lambda}^{3}M_{b}^{4}M_{w}^{2} +\qty(r+\sqrt{8 A_\lambda M_{b}^{2}+r^2})^6}{8 \sqrt{8 A_\lambda M_{b}^{2}+r^2}\qty(r+\sqrt{8 A_\lambda M_{b}^{2}+r^2})^3}, \\ \tilde{a} &=& \frac{1}{b^2(r)}\qty(1+\frac{r^2}{8 A_\lambda M_{b}^{2}}) \qty(1-\frac{2M_b}{\sqrt{8 A_\lambda M_{b}^{2}+r^2}})\,.\ee
We will focus on the physically interesting case of the symmetric bounce in which  $M_b=M_w$.
We treat $A_\lambda$ as the deformation parameter and calculate the resulting modification to the binary inspiral phasing imposed by this parameter.
Note that the deformation parameter satisfies the following conditions
\be\label{eq.lqg.delta.condition} A_\lambda>0 \,.\ee

%Brahma:2020eos 

The modification to the binary inspiral phasing induced by the deformation parameter $A_\lambda$ can be expressed as
\be\label{eq.lqg.phase.correction} \Psi_{\rm GW}^{\rm NGR}(f)=-\frac{5}{8u\eta^{2/5}} A_\lambda   \,,\ee
where $u=\eta^{3/5}\pi mf$.
Utilizing the ppE parameterization, this phase correction \eqref{eq.lqg.phase.correction} can be expressed as\be \label{eq.lqg.beta.exp}\beta= -\frac{5}{8\eta^{2/5}} A_\lambda \,.\ee
The modification enters at $b=-1$, corresponding to the $i=2$ and $1$PN order. 
We thus find that 
\be\label{eq.lqg.phase.1} A_\lambda= -\frac{3}{80}\varphi_2\delta\varphi_2\,.\ee
%=0.02^{+0.06}_{-0.06}

We take into account the natural restriction on the deformation parameter given by \eqref{eq.lqg.delta.condition}. 
The constraint is $A_\lambda=0.02^{+0.06}_{-0.02}$, implying that $A_\lambda<0.08$. This result is consistent with GR within the 90\% confidence interval, showing no significant deviation.

From the EHT image of Sagittarius $\rm A^*$, the constraints are $A_\lambda < 0.0821$ ($1\sigma$) and $A_\lambda < 0.1834$ ($2\sigma$)~\cite{Afrin:2022ztr}. We are not aware of constraints on $A_\lambda$ from X-ray reflection spectroscopy reported in the literature.

\subsection{Non-commutative geometry: Lorentzian mass distribution}
\label{sec.NCG}

In the phenomenological domain inspired by non-commutative geometry (NCG), one situation we consider here is the extension of point-like gravitational sources through Lorentz distributions. An example studied in this context is the non-rotating Lorentzian non-commutative metric, which can be written as \cite{Nozari:2008rc,Anacleto:2019tdj}
\be g_{tt}= -\qty[1-\frac{2m(r)}{r}]  \,,g_{rr}=\qty(-g_{tt})^{-1}  \,,g_{\theta\theta}=r^2\,,g_{\varphi\varphi}=r^2\sin^2\theta \,,\ee where\be m(r)=\frac{2m}{\pi}\qty(\arctan\frac{r}{\sqrt{\pi\vartheta }m}-\frac{r\sqrt{\pi\vartheta}m}{r^2+\pi\vartheta m^2}) \,.\ee
We treat $\vartheta$ of the NCG BH as a deformation parameter and calculate the resulting modification to the binary inspiral phasing imposed by this parameter. $\vartheta$ is a quantity that reflects the characteristics of the mass distribution. It is typically assumed to be positive, that is, \be\label{eq.ncld.delta.condition} \vartheta>0  \,.\ee

Using the Newman-Janis transformation, we can obtain the Lorentzian NCG metric in Boyer-Lindquist coordinates \{$t,r,\theta,\varphi$\}, which can be written as~\cite{AraujoFilho:2024rss}
\be\label{eq.le.ncld} \dd s^2  &=& - \frac{\Delta -a^2\sin^2\theta}{\rho^2} \dd t^2+ \frac{\rho^2}{\Delta} \dd r^2+ \rho^2 \dd\theta^2\nonumber \\ &&  +\frac{\qty(r^2+a^2)^2-\Delta a^2\sin^2\theta}{\rho^2}\sin^2\theta \dd\varphi^2-\frac{2a\qty(r^2+a^2-\Delta) \sin^2\theta}{\rho^2} \dd t \dd\varphi\, , \ee 
where
\be  \rho^2 = r^2 + a^2 \cos^2\theta\qand \Delta = r^2 - 2m(r)r + a^2\,. \ee

The modification to the binary inspiral phasing induced by the deformation parameter $\vartheta$ can be expressed as
\be\label{eq.ncld.phase.correction} \Psi_{\rm GW}^{\rm NGR}(f)= -\frac{5}{6\sqrt\pi\eta^{2/5}u}\sqrt{\vartheta} \,,\ee
where $u=\eta^{3/5}\pi mf$.
Utilizing the ppE parameterization, we find
\be \label{eq.ncld.beta.exp}\beta=-\frac{5}{6\sqrt\pi\eta^{2/5}}\sqrt{\vartheta} \,.\ee
The modification enters at $b=-1$, corresponding to the $i=2$ and $1$PN order. 
To facilitate comparison with GW observations, we have 
\be\label{eq.ncld.phase.1} \sqrt{\vartheta}=-\frac{9\sqrt\pi}{320} \varphi_2\delta\varphi_2\,.\ee
%=0.03^{+0.08}_{-0.08}

As for the other BH metrics, we use the GW170608 data with the SEOBNRv4P waveform model to constrain the deformation parameter. The constraint at the 90\% confidence level is $\sqrt\vartheta=0.03^{+0.08}_{-0.03}$, implying that $\sqrt\vartheta<0.11$. This result is consistent with GR within the 90\% confidence interval, showing no significant deviation.

From the EHT image of Sagittarius $\rm A^*$, the constraints on $\vartheta$ are $\vartheta<0.02$ ($1\sigma$) and $\vartheta<0.04$ ($2\sigma$)~\cite{Vagnozzi:2022moj}. We are not aware of any publication in the literature constraining $\vartheta$ with X-ray reflection spectroscopy.

\subsection{Kerr-Sen black hole}
\label{sec.KS}

The Sen BH is a non-regular space-time appearing in the low-energy limit of heterotic string theory~\cite{Gross:1984dd}.
The static spherically Sen BH can be written as~\cite{Sen:1992ua} \be -g_{tt}=(g_{rr})^{-1}=1-\frac{2M}{r+q^2/M} \,,g_{\theta\theta}=r^2\,,g_{\varphi\varphi}=r^2\sin^2\theta \,.\ee
By utilizing the string target space duality rotation, Sen discovered the rotating Sen (Kerr-Sen) BH~\cite{Sen:1992ua}. Alternatively, the same solution can be obtained through the Newman-Janis method~\cite{Yazadjiev:1999ce}.
The metric of the Kerr-Sen BH in Boyer-Lindquist coordinates \{$t,r,\theta,\varphi$\} can be written as
\be\label{eq.le.ks} \dd s^2 = -\qty(1-\frac{2mr}{\rho^2})\dd t^2+ \frac{\rho^2}{\Delta} \dd r^2+ \rho^2 \dd\theta^2  +\frac{\qty[r(r+r_\alpha)+a^2]^2-\Delta a^2\sin^2\theta}{\rho^2}\sin^2\theta \dd\varphi^2-\frac{4amr\sin^2\theta}{\rho^2} \dd t \dd\varphi\, ,\ee
where \be \Delta = r(r + r_\alpha m) - 2mr + a^2 \qand \rho^2 = r(r + r_\alpha m) + a^2 \cos^2\theta\,. \ee
We treat the $r_\alpha$ of the Kerr-Sen BH as the deformation parameter and calculate the resulting modification to the binary inspiral phasing imposed by this parameter. Here, $r_\alpha$ represents a quantity related to the effective charge, which encompasses both the electric charge and an effective charge associated with the dilaton field, as measured by a static observer at infinity. This quantity characterizes the specific hair and is constrained by the condition that  \be\label{eq.ks.delta.condition}  0<r_\alpha<2   \,.\ee

The modification to the binary inspiral phasing induced by the deformation parameter $r_\alpha$ can be expressed as
\be\label{eq.ks.phase.correction} \Psi_{\rm GW}^{\rm NGR}(f)=-\frac{405 r_\alpha}{128u^{1/3}\eta^{4/5}} \,,\ee
where $u=\eta^{3/5}\pi mf$.
Utilizing the ppE parameterization, we find
\be \label{eq.ks.beta.exp}\beta= -\frac{405}{128\eta^{4/5}}  r_\alpha \,.\ee
The modification enters at $b=-1/3$, corresponding to the $i=4$ and $2$PN order. 
We can also write
\be\label{eq.ks.phase.1}  r_\alpha=-\frac{1}{135} \varphi_4\delta\varphi_4\,.\ee

The constraint from the GW170608 data with the SEOBNRv4P waveform model is $r_\alpha=0.13^{+0.45}_{-0.13}$,  implying that $r_\alpha<0.58$, and it is consistent with the Kerr metric.

From the EHT image of Sagittarius $\rm A^*$, the constraints are $q/m<0.6$ ($1\sigma$) and $q/m<0.75$ ($2\sigma$)~\cite{Vagnozzi:2022moj}, where the relationship between the deformation parameter employed in Ref.~\cite{Vagnozzi:2022moj} and that in our study is $r_\alpha=q^2/m^2$. In the case of X-ray reflection spectroscopy, the parameter $r_\alpha$ was constrained from the analysis of a NuSTAR spectrum of the BH binary EXO 1846-031: $r_\alpha < 0.011$ (90\% confidence level)~\cite{Tripathi:2021rwb}.

\subsection{Ghosh-Kumar black hole}
\label{sec.Ghosh-Kumar}

An intriguing BH solution, referred to as the Ghosh-Kumar BH, was first proposed in Ref.~\cite{Ghosh:2021clx}. Although the metric functions discussed above remain regular throughout the entire spacetime, a scalar polynomial singularity persists at $r = 0$.
The metric of Ghosh-Kumar BH in Boyer-Lindquist coordinates \{$t,r,\theta,\varphi$\} can be written as~\cite{Ghosh:2021clx} 
\be\label{eq.le.gk} \dd s^2  &=& - \frac{\Delta -a^2\sin^2\theta}{\rho^2} \dd t^2+ \frac{\rho^2}{\Delta} \dd r^2+ \rho^2 \dd\theta^2\nonumber \\ &&  +\frac{\qty(r^2+a^2)^2-\Delta a^2\sin^2\theta}{\rho^2}\sin^2\theta \dd\varphi^2-\frac{2a\qty(r^2+a^2-\Delta) \sin^2\theta}{\rho^2} \dd t \dd\varphi\, , \ee 
where
\be  \rho^2 = r^2 + a^2 \cos^2\theta\qand \Delta = r^2 - 2m(r)r + a^2\,, \ee
and \be m(r)=\frac{mr}{\sqrt{r^2 + g^2 m^2}}\,.\ee
We treat the parameter $g$ of the Ghosh-Kumar BH as the deformation parameter of this metric and we calculate the resulting modification to the binary inspiral phasing imposed by this parameter.  As in reference~\cite{Vagnozzi:2022moj}, we focus on the case where\be\label{eq.gk.delta.condition} g>0  \,.\ee

The modification to the binary inspiral phasing induced by the deformation parameter $g$ can be expressed as
\be\label{eq.gk.phase.correction}\Psi_{\rm GW}^{\rm NGR}(f)=-\frac{75{ g}^2}{64u^{1/3}\eta^{4/5}} \,,\ee
where $u=\eta^{3/5}\pi mf$.
Utilizing the ppE parameterization, we get
\be \label{eq.gk.beta.exp}\beta= -\frac{75}{64\eta^{4/5}} { g}^2\,.\ee
The modification enters at $b=-1/3$, corresponding to the $i=4$ and $2$PN order and 
\be\label{eq.gk.phase.1} {g}^2=-\frac{1}{50} \varphi_4\delta\varphi_4\,.\ee
%= 0.36^{+1.20}_{-1.12}

The constraint from GW170608 is  $g^2=0.36^{+1.20}_{-0.36}$,  implying that $g^2<1.56$ and is consistent with GR.

From the EHT image of Sagittarius $\rm A^*$, the constraints are $g<1.4$ ($1\sigma$) and $g<1.6$ ($2\sigma$)~\cite{Vagnozzi:2022moj}. There are no constraints reported in the literature from X-ray reflection spectroscopy.

\subsection{Rotating Kazakov-Solodukhin black hole}
\label{sec.Rotating Kazakov-Solodukhin}

As an additional example of a regular BH, we consider the Kazakov-Solodukhin (KS) BH, which emerges within a string-inspired model where spherically symmetric quantum fluctuations of the metric and matter fields are described by the 2D dilaton-gravity action~\cite{Kazakov:1993ha}. 
The KS BH serves as a well-motivated example of the quantum deformation of the Schwarzschild BH. The metric function for this spacetime is expressed as
\be g_{tt}=-\qty(\sqrt{1 - \frac{b^2}{r^2}} - \frac{2M}{r})\,, g_{rr}=\qty(-g_{tt})^{-1}\,, g_{\theta\theta}=r^2\,,g_{\varphi\varphi}=r^2\sin^2\theta \,.\ee
We treat the parameter $b$ of the KS BH as a deformation parameter and calculate the resulting modification to the binary inspiral phasing imposed by this parameter.
In principle, $b$ can assume any positive value \be\label{eq.rks.delta.condition} b>0  \,,\ee
 as demonstrated in~\cite{Kocherlakota:2020kyu}.

Using the Newman-Janis transformation, we can obtain the rotating Kazakov-Solodukhin BH in Boyer-Lindquist coordinates \{$t,r,\theta,\varphi$\}, which can be written as~\cite{Xu:2021lff,Jusufi:2022tcw}
\be \label{eq.le.rks} \dd s^2  &=& - \frac{\Delta -a^2\sin^2\theta}{\rho^2} \dd t^2+ \frac{\rho^2}{\Delta} \dd r^2+ \rho^2 \dd\theta^2\nonumber \\ &&  +\frac{\qty(r^2+a^2)^2-\Delta a^2\sin^2\theta}{\rho^2}\sin^2\theta \dd\varphi^2-\frac{2a\qty(r^2+a^2-\Delta) \sin^2\theta}{\rho^2} \dd t \dd\varphi\, , \ee 
where
\be \rho^2 = r^2 + a^2 \cos^2\theta\qand \Delta = r^2 - 2m(r)r + a^2\,, \ee and \be  m(r)=m+\frac{r}{2}\qty(1-\sqrt{1-\frac{b^2m^2}{r^2}})\,.\ee

The modification to the binary inspiral phasing induced by the deformation parameter $b$ can be expressed as
\be\label{eq.rks.phase.correction} \Psi_{\rm GW}^{\rm NGR}(f)=\frac{5{b}^2}{96u\eta^{2/5}} \,,\ee
where $u=\eta^{3/5}\pi mf$.
Utilizing the ppE parameterization
\be \label{eq.rks.beta.exp}\beta= \frac{5}{96\eta^{2/5}} {b}^2 \,.\ee
The modification enters at $b=-1$, corresponding to the $i=2$ and $1$PN order. 
To facilitate comparison with GW observations, we write 
\be\label{eq.rks.phase.1} {b}^2=\frac{9}{20} \varphi_2\delta\varphi_2\,.\ee
%=-0.27^{+0.69}_{-0.71}

Using the GW170608 data with the SEOBNRv4P waveform model, we find $b^2=0.0^{+0.42}_{-0.0}$, implying that $b^2<0.42$.

From the EHT image of Sagittarius $\rm A^*$, the constraints are $l<0.2$ ($1\sigma$) and $l<1$ ($2\sigma$)~\cite{Vagnozzi:2022moj}. There are no studies reported in the literature to constrain $l$ with X-ray reflection spectroscopy.

\subsection{Ghosh regular black hole}
\label{sec.Ghosh}

The Ghosh regular BH is introduced in a purely phenomenological framework. The metric utilizes mass distribution functions that are inspired by continuous probability distributions. The metric of this BH can be written as~\cite{Balart:2014cga,Culetu:2014lca}
\be\label{eq.nonrotating.exp} g_{tt}= -\qty(1-\frac{2m(r)}{r})  \,,g_{rr}=\qty(-g_{tt})^{-1}  \,,g_{\theta\theta}=r^2\,,g_{\varphi\varphi}=r^2\sin^2\theta \,,\ee where \be m(r)=me^{-hm/r} \,.\ee
We treat $h$ as its deformation parameter and calculate the resulting modification to the binary inspiral phasing imposed by this parameter. As in reference~\cite{Vagnozzi:2022moj}, we focus on the case where
\be\label{eq.grbh.delta.condition} h>0  \,.\ee

By applying the Newman-Janis transformation, we can derive the rotating version of \eqref{eq.nonrotating.exp}, which represents the Ghosh regular BH. The metric of the Ghosh regular BH in Boyer-Lindquist coordinates \{$t,r,\theta,\varphi$\} can be written as~\cite{Ghosh:2014pba}
\be \label{eq.le.grbh}\dd s^2  &=& - \frac{\Delta -a^2\sin^2\theta}{\rho^2} \dd t^2+ \frac{\rho^2}{\Delta} \dd r^2+ \rho^2 \dd\theta^2\nonumber \\ &&  +\frac{\qty(r^2+a^2)^2-\Delta a^2\sin^2\theta}{\rho^2}\sin^2\theta \dd\varphi^2-\frac{2a\qty(r^2+a^2-\Delta) \sin^2\theta}{\rho^2} \dd t \dd\varphi\, , \ee 
where
\be  \rho^2 = r^2 + a^2 \cos^2\theta\qand \Delta = r^2 - 2m(r)r + a^2\,, \ee

The modification to the binary inspiral phasing induced by the deformation parameter $h$ can be expressed as
\be\label{eq.grbh.phase.correction} \Psi_{\rm GW}^{\rm NGR}(f)=-\frac{5{ h}}{24u\eta^{2/5}}  \,,\ee
where $u=\eta^{3/5}\pi mf$.
Utilizing the ppE parameterization
\be \label{eq.grbh.beta.exp} \beta= -\frac{5}{24\eta^{2/5}}{ h} \,.\ee
The modification enters at $b=-1$, corresponding to the $i=2$ and $1$PN order. 
To facilitate comparison with GW observations, we write 
\be\label{eq.grbh.phase.1} {  h}=-\frac{9}{80} \varphi_2\delta\varphi_2\,.\ee
%=0.07^{+0.18}_{-0.17}

From the GW170608 data with the SEOBNRv4P waveform model, the constraint is $h=0.07^{+0.18}_{-0.07}$, implying that $h<0.25$, and it is consistent with GR.

From the EHT image of Sagittarius $\rm A^*$, the constraints are $g/m<0.8$ ($1\sigma$) and $g<m$ ($2\sigma$) respectively~\cite{Vagnozzi:2022moj}. The relationship between the parameter used in Ref.~\cite{Vagnozzi:2022moj} and in our study is $h=g^2/2m^2$. We are not aware of any study testing the Ghosh regular BH metric with X-ray reflection spectroscopy.

\subsection{Tinchev black hole}
\label{sec.Tinchev}

Similar to the Ghosh regular BH, the Tinchev BHcan be obtained by modifying the form of the mass function.
The metric of the Tinchev BH in Boyer-Lindquist coordinates \{$t,r,\theta,\varphi$\} can be written as~\cite{Tinchev:2015apf}
\be \label{eq.le.tbh}\dd s^2  &=& - \frac{\Delta -a^2\sin^2\theta}{\rho^2} \dd t^2+ \frac{\rho^2}{\Delta} \dd r^2+ \rho^2 \dd\theta^2\nonumber \\ &&  +\frac{\qty(r^2+a^2)^2-\Delta a^2\sin^2\theta}{\rho^2}\sin^2\theta \dd\varphi^2-\frac{2a\qty(r^2+a^2-\Delta) \sin^2\theta}{\rho^2} \dd t \dd\varphi\, , \ee 
where
\be  \rho^2 = r^2 + a^2 \cos^2\theta\qand \Delta = r^2 - 2m(r)r + a^2\,, \ee
and\be m(r)=m e^{-jm^2/r^2}    \,.\ee
We treat $j$ as the deformation parameter and we calculate the resulting modification to the binary inspiral phasing imposed by this parameter. The deformation parameter satisfies
\be\label{eq.tbh.delta.condition} j>0 \,.\ee

The modification to the binary inspiral phasing induced by the deformation parameter $j$ can be expressed as
\be\label{eq.tbh.phase.correction} \Psi_{\rm GW}^{\rm NGR}(f)=-\frac{75{ j}}{32u^{1/3}\eta^{4/5}} \,,\ee
where  $u=\eta^{3/5}\pi mf$.
Utilizing the ppE parameterization
\be \label{eq.tbh.beta.exp}\beta=-\frac{75{ j}}{32\eta^{4/5}} \,.\ee
The modification enters at $b=-1/3$, corresponding to the $i=4$ and $2$PN order. 
We can also write 
\be\label{eq.tbh.phase.1}  j=-\frac{1}{100} \varphi_4\delta\varphi_4\,.\ee
%tbh= 0.18^{+0.60}_{-0.56}

From the GW170608 data, the constraint is $j= 0.18^{+0.60}_{-0.18}$, implying that $j<0.78$.

We are not aware of constraints on $j$ reported in the literature with BH shadows or X-ray reflection spectroscopy.

\subsection{Rotating Einstein-Born-Infeld black hole}
\label{Rotating EBI}

The gravitational field of a static, spherically symmetric compact object with mass $m$ and a nonlinear electromagnetic source within the framework of the Einstein-Born-Infeld theory was first investigated by Hoffmann~\cite{Hoffmann:1935ty}. The corresponding spacetime metric is given by~\cite{Hoffmann:1935ty,Gibbons:1995cv,Gibbons:1996pd,Chruscinski:2000zm,Sorokin:1997nz}:
\be g_{tt}= -\qty(1-\frac{2m(r)}{r})  \,,g_{rr}=\qty(-g_{tt})^{-1}  \,,g_{\theta\theta}=r^2\,,g_{\varphi\varphi}=r^2\sin^2\theta   \,,\ee
and
\be m(r)=m-\frac{Q^2(r)}{2r}\,, \ee
where
\be Q^2(r) = \frac{2\beta^2 r^4}{3} \qty(1 - \sqrt{1 + \zeta^2(r)})+ \frac{4Q^2m^2}{3} F\qty(\frac{1}{4}, \frac{1}{2}, \frac{5}{4}, -\zeta^2(r)) %\simeq Q^2m^2-\frac{Q^4m^4}{20r^4\beta^2}+\order{Q^6,\beta^{-4}}
\,. \ee $F$ is a Gauss hypergeometric function and $\zeta^2(r)=Q^2m^2/\beta^2r^4$.
$Q$ is the deformation parameter and we calculate the resulting modification to the binary inspiral phasing imposed by this parameter.
We typically assume that the parameter $Q$ is positive, that is: \be\label{eq.ebi.delta.condition} Q>0  \,.\ee

By applying the Newman-Janis transformation, we can derive the rotating version of the Einstein-Born-Infeld BH.
The metric of the rotating Einstein-Born-Infeld BH in Boyer-Lindquist coordinates \{$t,r,\theta,\varphi$\} can be written as~\cite{CiriloLombardo:2004qw,Atamurotov:2015xfa}
\be\label{eq.le.ebi} \dd s^2  &=& - \frac{\Delta -a^2\sin^2\theta}{\rho^2} \dd t^2+ \frac{\rho^2}{\Delta} \dd r^2+ \rho^2 \dd\theta^2\nonumber \\ &&  +\frac{\qty(r^2+a^2)^2-\Delta a^2\sin^2\theta}{\rho^2}\sin^2\theta \dd\varphi^2-\frac{2a\qty(r^2+a^2-\Delta) \sin^2\theta}{\rho^2} \dd t \dd\varphi\, , \ee 
and
\be  \rho^2 = r^2 + a^2 \cos^2\theta\qand \Delta = r^2 - 2m(r)r + a^2\,. \ee

The modification to the binary inspiral phasing induced by the deformation parameter $Q$ can be expressed as
\be\label{eq.ebi.phase.correction} \Psi_{\rm GW}^{\rm NGR}(f)=-\frac{5 }{48 u \eta^{2/5}}Q^2  \,,\ee
where $u=\eta^{3/5}\pi mf$.
Utilizing the ppE parameterization
\be \label{eq.ebi.beta.exp}\beta=-\frac{5 }{48\eta^{2/5}}Q^2 \,.\ee
The modification enters at  $b=-1$, corresponding to the  $i=2$ and $1$PN order. 
To facilitate comparison with GW observations, we write 
\be\label{eq.ebi.phase.1} Q^2=-\frac{9}{40}\varphi_2\delta\varphi_2\,.\ee
%= 0.13^{+0.35}_{-0.34}

From the GW170608 data with the SEOBNRv4P waveform model, the constraint is $Q^2= 0.13^{+0.35}_{-0.13}$, implying that $Q^2<0.48$.

To date, no studies have been published that use BH shadows or X-ray reflection spectroscopy to constrain the deformation parameter of  this metric.

\subsection{Balart-Vagenas black holes}
\label{sec.other}

Similar to the metrics discussed in the previous subsections, various regular BH solutions can be constructed by employing different mass functions. In Ref.~\cite{Balart:2014cga}, several charged regular BH solutions are derived within the framework of theories that couple nonlinear electrodynamics with GR. The corresponding mass functions are obtained from continuous probability distributions, such as:
\begin{itemize}
	\item $m(r)=2m/(1+e^{q^2m/r})$\footnote{It is noteworthy that this metric corresponds to an Ay\'on-Beato and Garc\'ia BH~\cite{Ayon-Beato:1999kuh}.},
	\item $m(r)=4m e^{-\sqrt{2q^2m/r}}/(1+e^{-\sqrt{2q^2m/r}})^2$,
	\item $m(r)=m\qty(\cosh{\sqrt{\frac{q^2m}{r}}})^{-1}$,
	\item $m(r)=\frac{q^2m^2}{ r}/\qty[\exp(\frac{q^2m}{ r})-1]$,
	\item $m(r)=\frac{6q^2m^2}{ r}\exp(\sqrt{\frac{6q^2m}{r}})/\qty[\exp(\sqrt{\frac{6q^2m}{r}})-1]^2$.
\end{itemize}

We consider the line element for the static and spherically symmetric metric,
\be\label{eq.nonrotating.exp1} g_{tt}= -\qty(1-\frac{2m(r)}{r})  \,,g_{rr}=\qty(-g_{tt})^{-1}  \,,g_{\theta\theta}=r^2\,,g_{\varphi\varphi}=r^2\sin^2\theta   \,,\ee
where $m$ can be any of the mass functions mentioned above. The outer and the inner horizons are located at \(r_+\) and \(r_-\), respectively, satisfying \(r_{\pm} = 2m(r_{\pm})\).
The invariant scalars and electric fields of the regular BHs constructed in this way are regular throughout the entire spacetime.
We treat $q$ as the deformation parameter and we calculate the resulting modification to the binary inspiral phasing imposed by this parameter.
Here, we restrict the range of $q$ to \be\label{eq.regularBH.delta.condition} q>0  \,.\ee

By applying the Newman-Janis transformation, we can derive the rotating versions of \eqref{eq.nonrotating.exp1}.
These metrics in Boyer-Lindquist coordinates \{$t,r,\theta,\varphi$\} can be written as
\be\label{eq.le.regularBH} \dd s^2  &=& - \frac{\Delta -a^2\sin^2\theta}{\rho^2} \dd t^2+ \frac{\rho^2}{\Delta} \dd r^2+ \rho^2 \dd\theta^2\nonumber \\ &&  +\frac{\qty(r^2+a^2)^2-\Delta a^2\sin^2\theta}{\rho^2}\sin^2\theta \dd\varphi^2-\frac{2a\qty(r^2+a^2-\Delta) \sin^2\theta}{\rho^2} \dd t \dd\varphi\, , \ee 
where
\be  \rho^2 = r^2 + a^2 \cos^2\theta\qand \Delta = r^2 - 2m(r)r + a^2\,, \ee

The modification to the binary inspiral phasing induced by the deformation parameter $q$ can be expressed as
\be\label{eq.regularBH.phase.correction} \Psi_{\rm GW}^{\rm NGR}(f)=-\frac{5 }{48 u \eta^{2/5}}q^2  \,,\ee
where $u=\eta^{3/5}\pi mf$.
Utilizing the ppE parameterization 
\be \label{eq.regularBH.beta.exp}\beta=-\frac{5 }{48\eta^{2/5}}q^2 \,.\ee
The modification enters at  $b=-1$, corresponding to the  $i=2$ and $1$PN order. We can also write 
\be\label{eq.regularBH.phase.1} q^2=-\frac{9}{40}\varphi_2\delta\varphi_2\,.\ee
%= 0.13^{+0.35}_{-0.34}

When using GW170608 data with SEOBNRv4P waveform model to constrain the deformation parameter, we find $q^2= 0.13^{+0.35}_{-0.13}$, implying that $q^2<0.48$. 

To date, no studies have been found that use BH shadows or X-ray reflection spectroscopy to constrain the deformation parameter of  this metric.

\section{Brief discussion}
\label{sec:briefdiscussion}

We fit the publicly released posterior sample from the LIGO-Virgo  Collaboration~\cite{LVC-web} to obtain constraints on the deformation parameters of various BH spacetimes beyond GR.
We utilize the publicly available Markov Chain Monte Carlo (MCMC) samples for the best-fit model, as provided in the ``Tests of General Relativity with Binary Black Holes from the second LIGO-Virgo Gravitational-Wave Transient Catalog - Full Posterior Sample Data Release''\footnote{\href{https://zenodo.org/records/5172704}{https://zenodo.org/records/5172704}}.
For the event name, we adhere to the naming convention used in the scripts from the LIGO-Virgo Collaboration, as described in Ref.~\cite{LVC-web}

The procedure for obtaining constraints on the deformation parameters of various BH spacetimes from GW observations is straightforward, as demonstrated by the analysis above. Specifically, in our case, as indicated in Eq.~\eqref{eq.def.delta}, the deformation parameters depend on both $\varphi_i$ and $\delta\varphi_i$. The latter can be directly obtained from the LIGO-Virgo Collaboration, which analyzed the GW events to constrain all non-GR parameters. In contrast, $\varphi_i$ can be computed using \eqref{eq.i=0}, \eqref{eq.i=1}, \eqref{eq.i=2}, \eqref{eq.i=3}, and \eqref{eq.i=4}, where $\eta$, the symmetric mass ratio, depends on the BH masses $m_1$ and $m_2$, which are also included in the GW transient catalog-1~\cite{LIGOScientific:2018mvr}. It is important to emphasize that the deformation parameters for the various spacetimes considered here are positive. Therefore, when fitting these parameters, we exclude any negative values and only consider those that satisfy the condition $0 < \delta < \delta_{\text{extremal}}$.

\begin{figure}[H]
    \centering
    \subfigure{\includegraphics[width=0.3\textwidth]{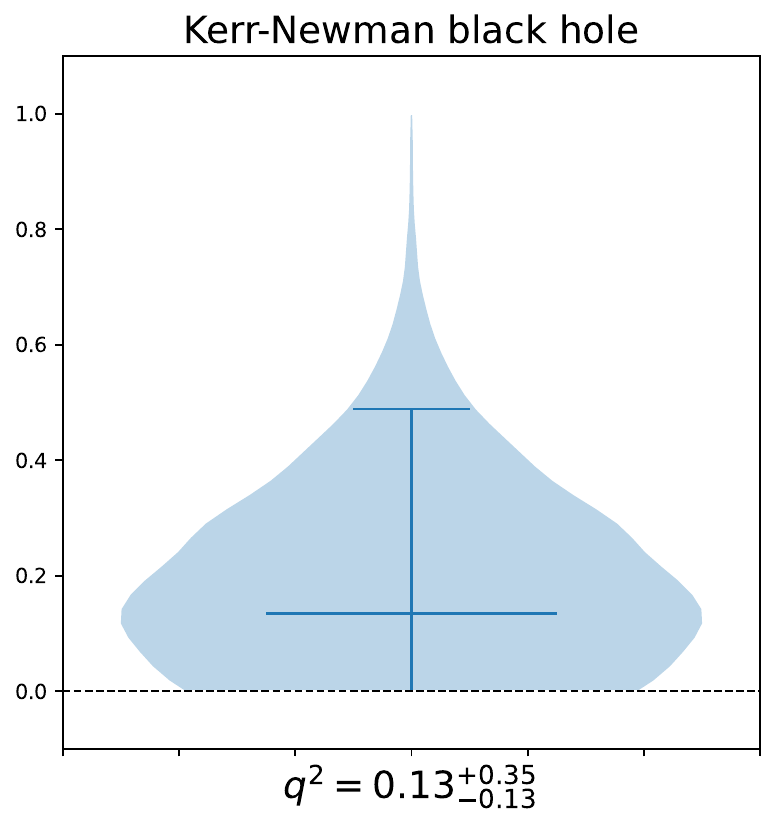}}\hspace{0.3em}
    \subfigure{\includegraphics[width=0.3\textwidth]{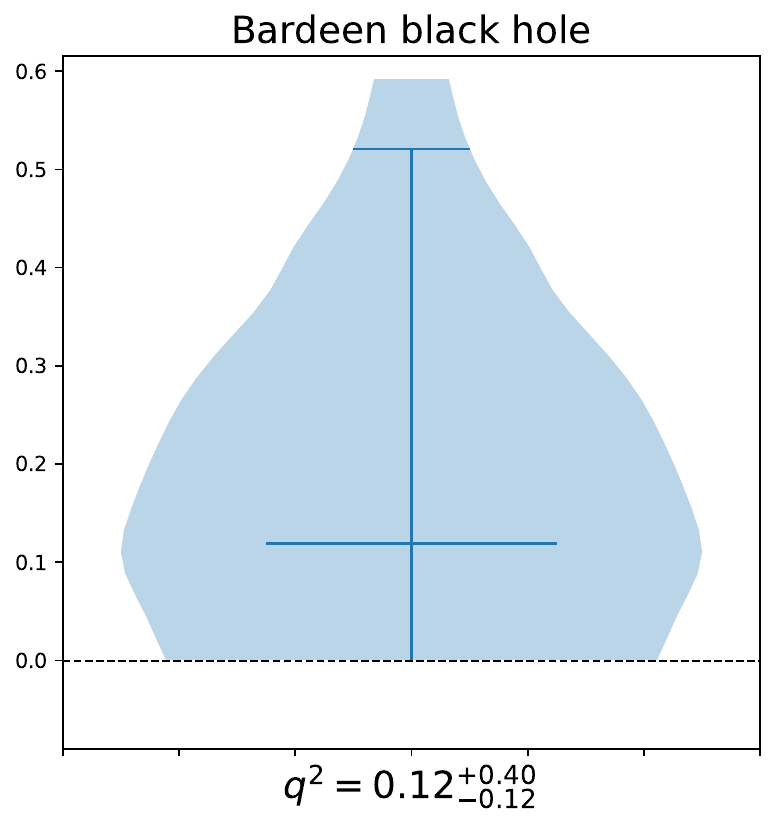}}\hspace{0.3em}
    \subfigure{\includegraphics[width=0.3\textwidth]{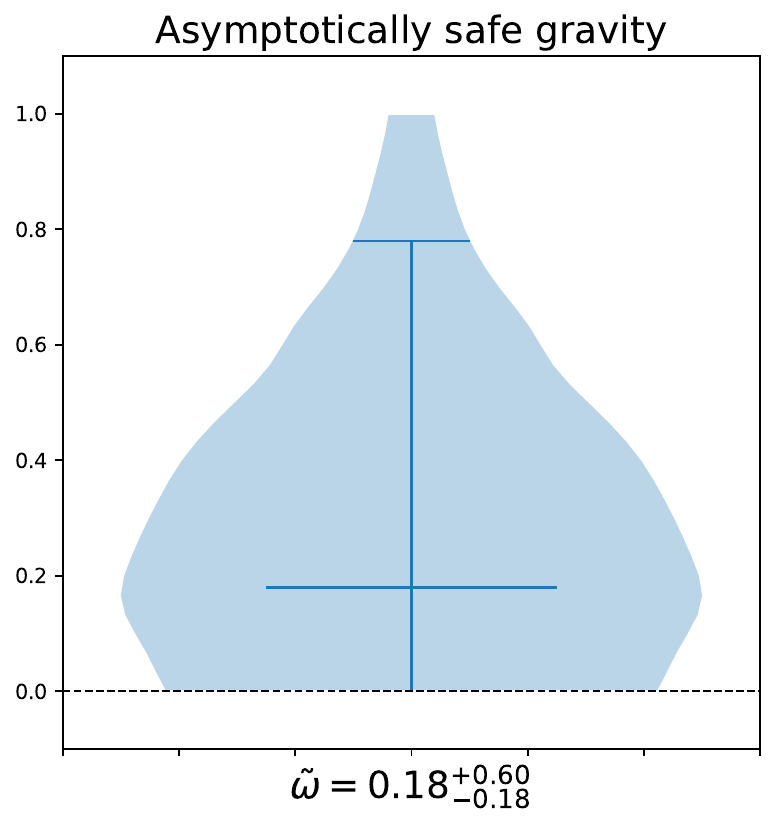}}\\[-1em]
    
    \subfigure{\includegraphics[width=0.3\textwidth]{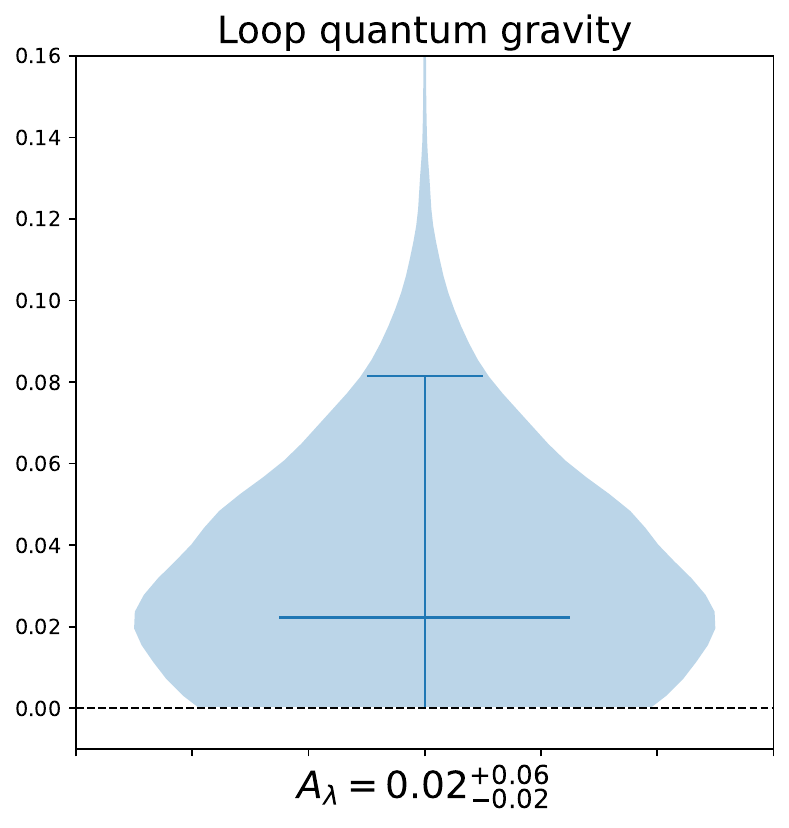}}\hspace{0.3em}
    \subfigure{\includegraphics[width=0.3\textwidth]{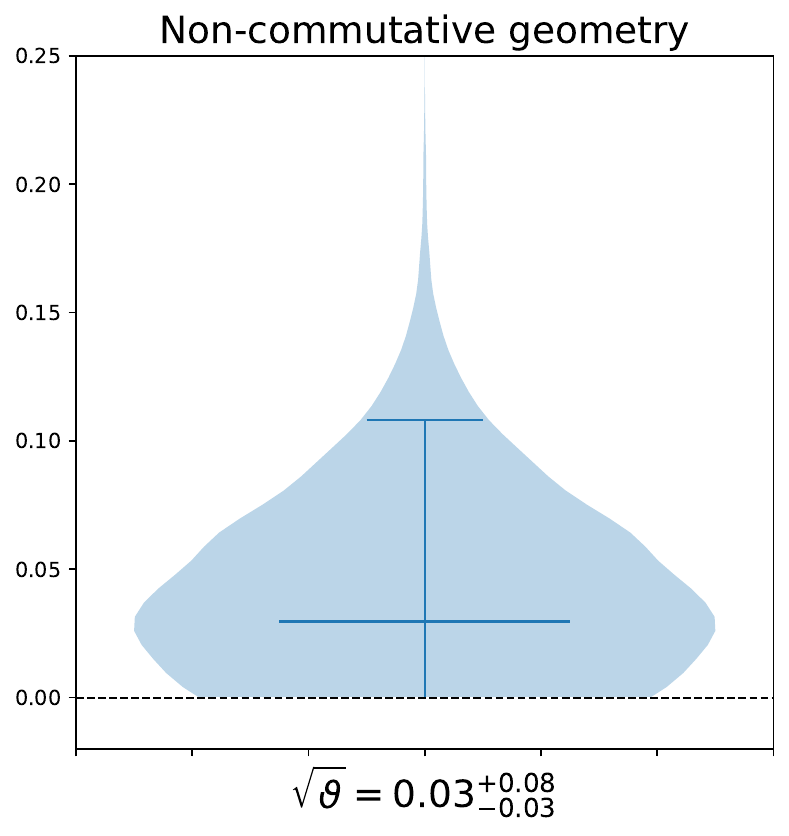}}\hspace{0.3em}
    \subfigure{\includegraphics[width=0.3\textwidth]{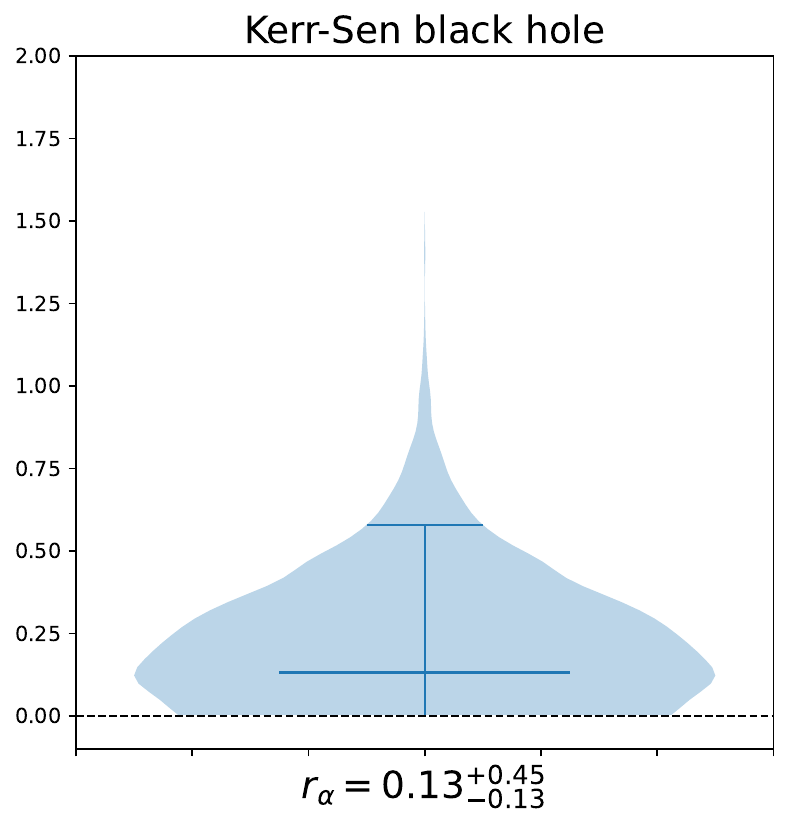}}\\[-1em]
   
    \subfigure{\includegraphics[width=0.3\textwidth]{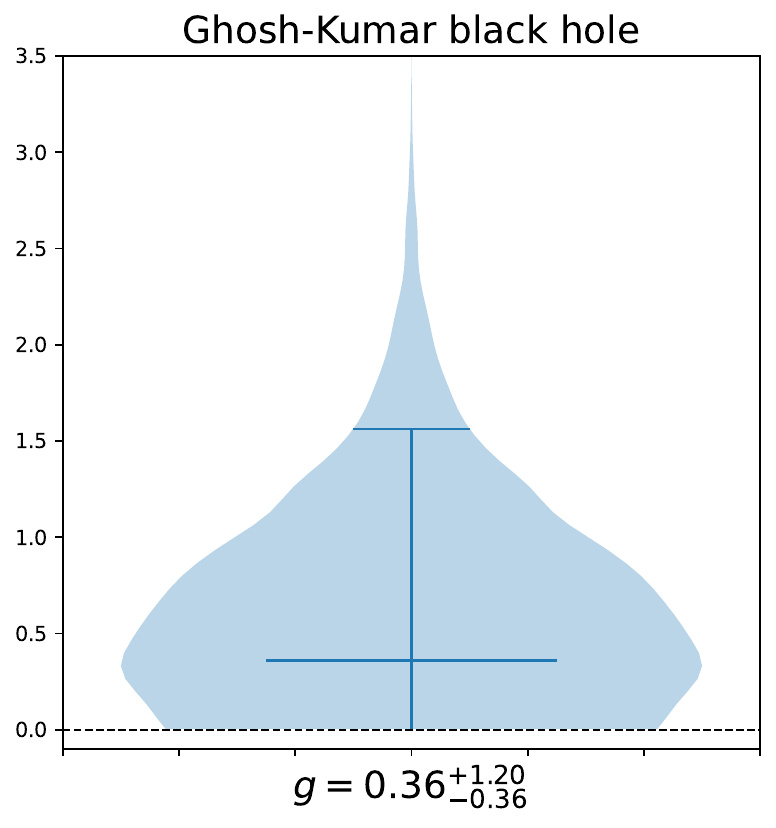}}\hspace{0.3em} 
    \subfigure{\includegraphics[width=0.3\textwidth]{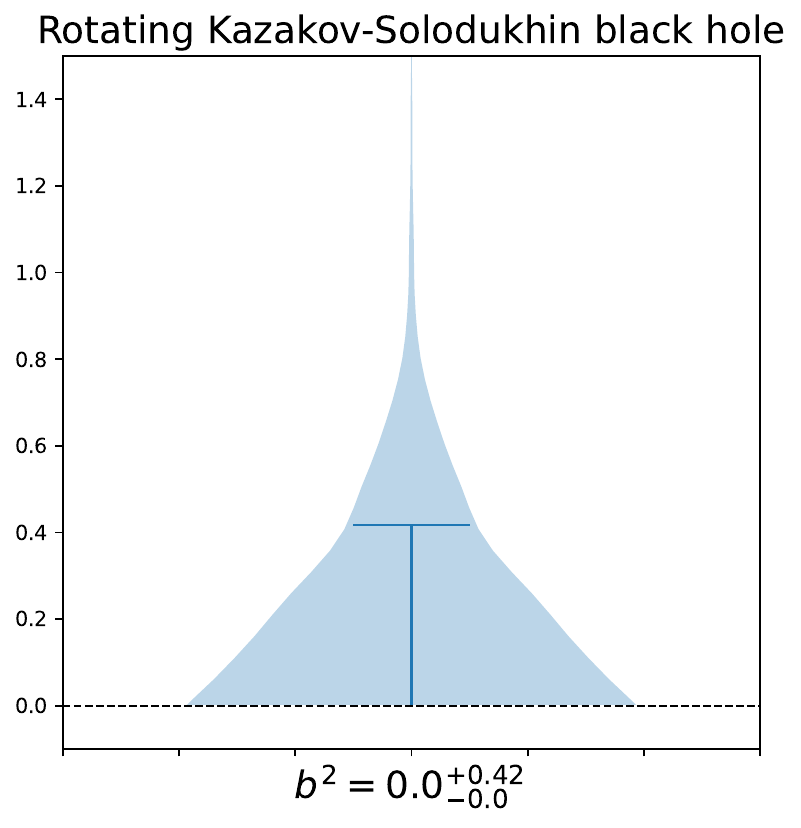}}\hspace{0.3em}
    \subfigure{\includegraphics[width=0.3\textwidth]{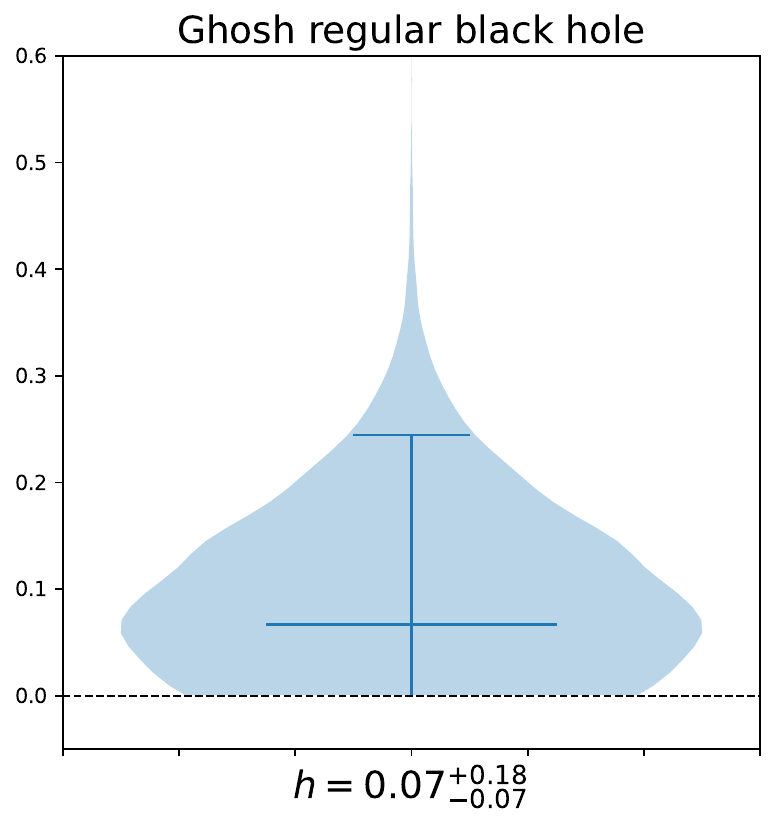}}\\[-1em]
    
    \subfigure{\includegraphics[width=0.3\textwidth]{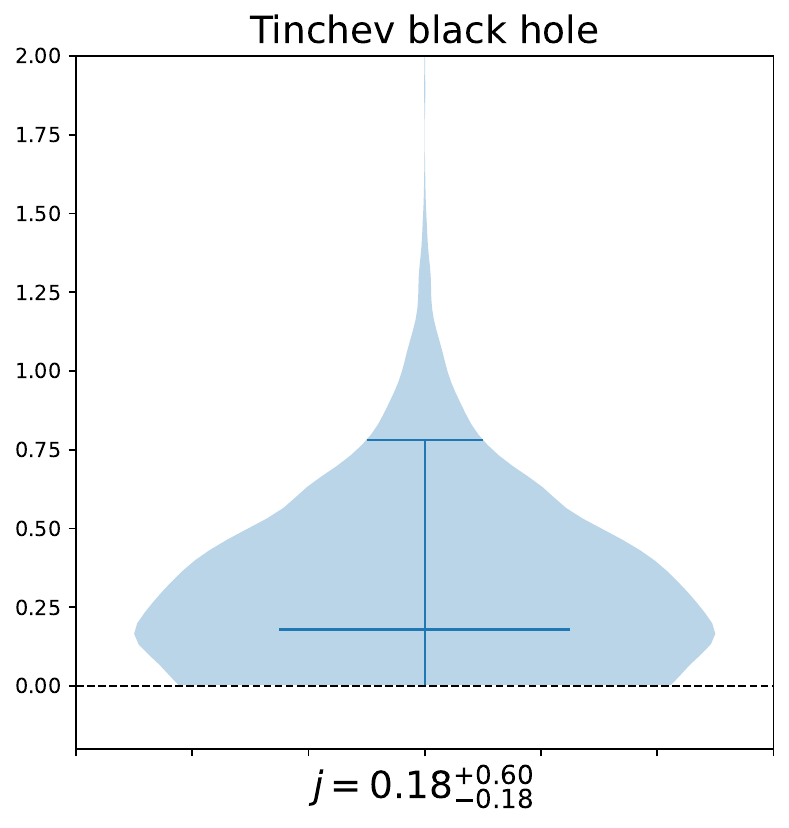}}\hspace{0.3em}
    \subfigure{\includegraphics[width=0.3\textwidth]{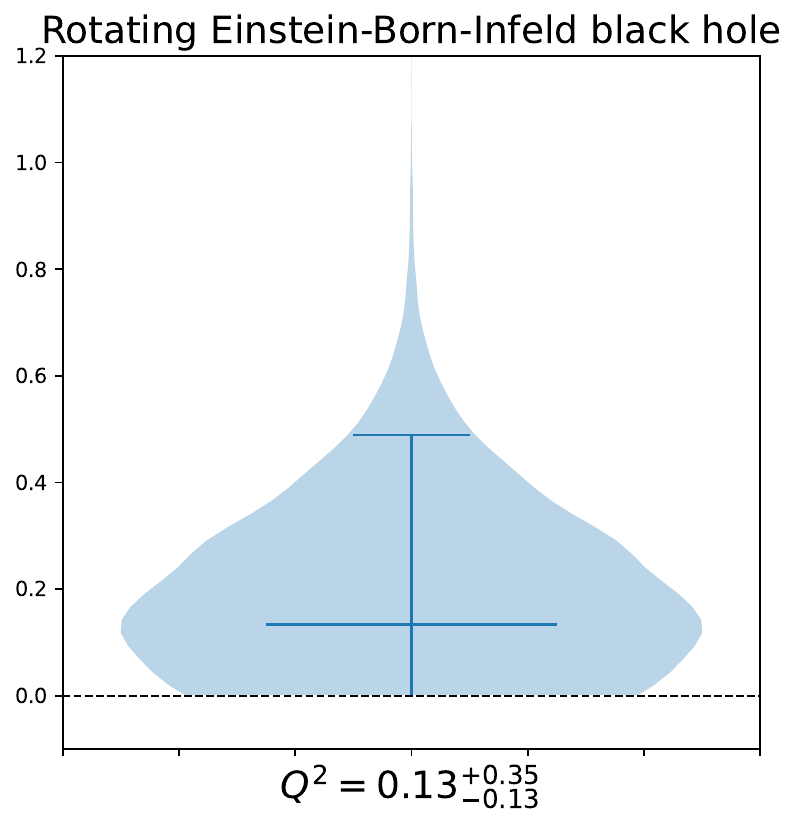}}   \hspace{0.3em}
    \subfigure{\includegraphics[width=0.3\textwidth]{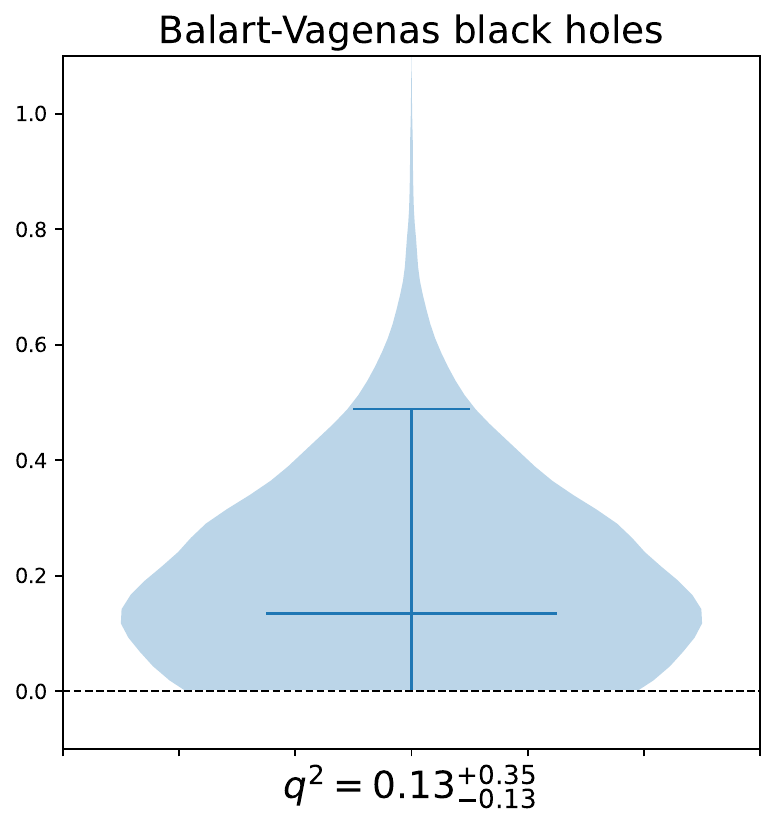}}
    
    \caption{Violin plots for the deformation parameters of the BH spacetimes beyond GR considered in this work. All constraints are inferred from the GW170608 data with the SEOBNRv4P waveform model. The thick horizontal line marks the median and the thin horizontal line marks the 90\% credible interval.}
    \label{fig.1}
\end{figure}

We utilized the GW170608 event to derive constraints on the deformation parameters across various models. This event involves a binary system with component masses of $\sim 12 M_\odot$ and $\sim 7 M_\odot$, respectively~\cite{LIGOScientific:2017vox}. 
The results are summarized in Table~\ref{table.11}, where we present the uncertainties at the 90\% confidence level.
Additionally, the table includes the constraints (if available in the literature) derived from BH shadow observations and X-ray reflection spectroscopy, allowing for a comparison of the constraints obtained through different techniques. 
Figure~\ref{fig.1} displays violin plots corresponding to the constraints from GW170608.

\begin{table}[H]
%\begin{table}[h]
	\centering
  \begin{threeparttable}[b]
    %\centering
   \begin{tabular}{cccc}
   \toprule
     BH metric         & GWs ($90\%$) & EHT & X-rays  \\ \midrule
   Kerr-Newman BH \S.\ref{sec.KN}    & $q^2<0.48$   & $q<0.8$, $0.95$ &    \\
   Bardeen BH   \S.\ref{sec.Bardeen}  &   $q^2<0.52$ & unconstrained  & $q < 0.41$ ($3\sigma$)$^\dag$  \\
   Asymptotically safe gravity \S.\ref{sec.Asymptoticallysafe}  & $\tilde\omega <0.78$ & $\tilde\omega<0.9$, unconstrained  & $\tilde{\omega} < 0.05$ (90\%) \\
  Loop quantum gravity \S.\ref{sec.lqg} & $A_\lambda<0.08$  & $A_\lambda < 0.0821$, $0.1834$ &   \\
  Non-commutative geometry  \S.\ref{sec.NCG}  &$\sqrt\vartheta<0.11$& $\vartheta<0.02$, $0.04$   \\
  Kerr-Sen black hole  \S.\ref{sec.KS}  &$r_\alpha<0.58$ & $\sqrt{r_\alpha}<0.6$, $0.75$  &$r_\alpha < 0.011$ (90\%) \\
  Ghosh-Kumar BH \S.\ref{sec.Ghosh-Kumar} &$g^2<1.56$ & $g<1.4$, $1.6$   \\
  Rotating Kazakov-Solodukhin BH \S.\ref{sec.Rotating Kazakov-Solodukhin} & $b^2<0.42$ & $l<0.2$, $1$   \\
  Ghosh regular BH  \S.\ref{sec.Ghosh}  & $h<0.25$  & $h<0.32$, $0.5$   \\
 Tinchev BH \S.\ref{sec.Tinchev}  &  $j<0.78$  &    \\
 Rotating Einstein-Born-Infeld BH \S.\ref{Rotating EBI}  &$Q^2<0.48$ &    \\
   Balart-Vagenas BHs \S.\ref{sec.other}  &$q^2<0.48$ &    \\
                    \bottomrule
    \end{tabular}
    \caption[table-fxyz]{\small Summary of the results. 
     The first column lists the names of the BH spacetimes beyond GR considered in our study and the corresponding subections. The second column provides a summary of the constraints on the deformation parameters. These constraints are derived using GW data of GW170608 with the SEOBNRv4P waveform model, at the 90\% confidence level. 
	The third column shows the $1\sigma$ and $2\sigma$ constraints on the deformation parameters obtained from BH shadow observations, as discussed in~\cite{Vagnozzi:2022moj}. The term ``unconstrained'' indicates that the parameter is not constrained because the shadow size remains consistent with EHT observations across all parameter values. Blank cells signify that no studies have been identified that constrain these models using BH shadow data. 
	The last column shows the constraints on the deformation parameters with X-ray reflection spectroscopy. Relevant findings are detailed in the references cited within the main text. Similarly, blank cells indicate the absence of corresponding studies.\\$^\dag$ This constraint on $q$ of the Bardeen metric was obtained from the thermal spectrum of the accretion disk of Cygnus~X-1, not with X-ray reflection spectroscopy.
    }
    \label{table.11}
  \end{threeparttable}
\end{table}

The above analysis is based on the assumption that the gravitational-wave signal is well described by quasi-circular templates. However, a natural question is to what extent unmodeled physical effects, such as orbital eccentricity, may influence the inferred deformation parameters. We therefore provide a brief assessment of the impact of eccentricity below.

\subsection{Impact of orbital eccentricity}

%Circular template, possibly influence of unmodeled eccentricity
The gravitational-wave event GW170608 was analyzed using quasi-circular waveform templates, such as SEOBNRv3~\cite{Pan:2013rra,Taracchini:2013rva,Babak:2016tgq} and IMRPhenomPv2~\cite{Hannam:2013oca,Husa:2015iqa,Khan:2015jqa}. If the true binary orbit possesses a small but nonzero eccentricity, the corresponding contribution to the waveform is not explicitly modeled and may therefore be partially absorbed into the inferred non-GR deviations.

%Constraint on the eccentricity of GW170608
Within GR, gravitational radiation tends to circularize binary orbits during the inspiral~\cite{Pfeiffer:2007yz}. Several studies have investigated the eccentricity of GW170608 and found it to be consistent with zero. In particular, Romero-Shaw et al.~\cite{Romero-Shaw:2019itr} obtained an upper bound of $e < 0.036$ at the $90\%$ confidence level, while O'Shea et al.~\cite{OShea:2021faf} reported $e < 0.12$ using TEOBResumE~\cite{Chiaramello:2020ehz}. A more conservative bound of $e < 0.166$ was obtained in \cite{Wu:2020zwr}, where the analysis was performed using the EccentricFD waveform model~\cite{Huerta:2014eca}. This model describes only the inspiral phase and does not include spin effects. Moreover, the eccentric contributions to the phase are not derived from a fully systematic PN expansion, but rather constructed via a heuristic extension of leading-order eccentric corrections to higher orders. The above results are all presented at the reference frequency of 10 Hz. It should be noted that eccentricity estimates may not be directly comparable across different waveform models~\cite{Knee:2022hth,Shaikh:2023ypz}.

%Some existing studies have shown that the eccentricity has little effect on some of the intrinsic parameters of BH.
Existing studies further indicate that eccentricity has a limited impact on the estimation of intrinsic parameters for GW170608-like systems. In particular, the inclusion of eccentricity does not lead to significant shifts in the recovered masses~\cite{OShea:2021faf}, and systematic biases are expected to become important only for relatively large eccentricities $e \gtrsim 0.1$--$0.2$~\cite{Favata:2021vhw}.

%ppE decomposition of phase
In the presence of small eccentricity and small deviations from GR, the Fourier-domain phase may be schematically written as
\be\label{eq.gwphase.tot1}
\Psi_{\rm GW}=\Psi_{\rm GR}^{\rm circ.}+\Psi_{\rm GR}^{\rm ecc.}+\Delta\Psi
=\Psi_{\rm GR}^{\rm circ.}+\Delta\Psi'\,,
\ee
where $\Delta\Psi'=\Delta\Psi+\Psi_{\rm GR}^{\rm ecc.}$ denotes the total deviation with respect to the circular GR baseline. This implies that the ppE parameters inferred from circular-template analyses may effectively include contributions from both modified-gravity effects and unmodeled eccentricity.
Within the ppE formalism, the gravitational-wave phase in Eq.~\eqref{eq.gwphase.tot1} is typically expressed in the explicit form
\be
\Psi_{\rm GW}\sim \frac{3}{128}u^{-5/3}\sum_i \varphi_i\qty(1+\delta \varphi_i) u^{i/3}\eta^{-i/5}
\sim \frac{3}{128}u^{-5/3}\sum_i \varphi_i^{\rm circ.}\qty(1+\delta \varphi_i') u^{i/3}\eta^{-i/5}\,,
\ee
where $\delta \varphi_i$ parametrizes possible deviations from GR, and the effective deviation parameter $\delta \varphi_i'$ is given by
\be\label{eq.eandpostgr}
\delta \varphi_i'=\varphi_i^{\rm ecc.}+\delta \varphi_i\,.
\ee
This expression makes it clear that, when a circular waveform template is employed to fit the signal, part of the inferred ppE deviation parameters can, in fact, be attributed to residual eccentricity effects.

%equation of ecc
To assess the impact of orbital eccentricity on the estimation of deformation parameters, it is necessary to compare the magnitude of $\varphi_i^{\rm ecc.}$ with that of $\delta \varphi_i'$ in Eq.~\eqref{eq.eandpostgr}. In fact, $\varphi_i^{\rm ecc.}$ represents the eccentricity-dependent contributions to the PN phase coefficients. The leading-order corrections to the PN coefficients induced by eccentricity at $0$PN order were first derived in \cite{Krolak:1995md,Yunes:2009yz}, and subsequently extended to $2$PN and $3$PN orders in \cite{Tanay:2016zog} and \cite{Moore:2016qxz}, respectively.
For instance, at $1$PN and $2$PN orders, the eccentricity-dependent contributions to the PN coefficients can be expressed as
\be \label{eq.PNecc.1}
\varphi_2^{\rm ecc.}=\qty{ \qty(-\frac{2045665}{348096}-\frac{128365}{12432}\eta)\chi^{-19/9} + \qty(-\frac{2223905}{491232}+\frac{154645}{17544}\eta )\chi^{-25/9}}e_0^{2} +\order{e_0^{4}}  \,,
\ee
and
\be \label{eq.PNecc.2}
\varphi_4^{\rm ecc.}= \left\lbrace  \qty(- \frac{111064865}{ 14 141 952}-  \frac{165068815}{ 4124736 }\eta - \frac{10 688 155}{294624}\eta^2)\chi^{-19/9} +\left( -\frac{5 795 368 945}{350 880 768} + \frac{4 917 245 }{ 1 566 432}\eta \right.\right.\nonumber\\
\left.\left.{} + \frac{25 287 905}{447552}\eta^2  \right)\chi^{-25/9} +  \qty( \frac{936702035}{1 485 485 568 }+ \frac{3062285}{ 260064 }\eta - \frac{14 251 675 }{631 584}\eta^2)\chi^{-31/9} \right\rbrace   e_0^2 +\order{e_0^{4}} \,,
\ee 
where $\chi=f/f_0$, and $f_0$ is the reference frequency we have chosen, i.e., $e_t(f_0)=e_0$. 
These expressions explicitly show how eccentricity enters the PN phase through corrections proportional to $e_0^2$, with higher-order contributions suppressed by $\order{e_0^4}$.

%estimation
Assuming that the intrinsic parameters are not significantly affected by eccentricity, one may estimate its impact on deformation parameters by comparing the magnitude of the eccentric phase corrections with that of the inferred deviations.
For GW170608, the parameter-estimation analysis adopts different low-frequency cutoffs for the two detectors, with $f_{\rm low}=30\,\mathrm{Hz}$ for LHO and $f_{\rm low}=20\,\mathrm{Hz}$ for LLO, as reported in the LIGO/Virgo analysis. Since the eccentricity-induced phase corrections scale with negative powers of frequency (e.g., $f^{-19/9}$), their contribution is dominated by the low-frequency regime.
To obtain a conservative estimate of the maximal impact of eccentricity, we therefore evaluate the phase at a frequency of $20\,\mathrm{Hz}$, corresponding to the lowest frequency used in the analysis. Additionally, the eccentricity value chosen is $e_0 = 0.1$. We then obtain
\be
\varphi_2^{\rm ecc.}=-0.0229^{+0.0007}_{-0.0001}\,, \quad \varphi_4^{\rm ecc.}=-0.0617^{+0.0049}_{-0.0007}\,,
\ee
which can be compared with the reported constraints
\be
\delta\varphi_2'= -0.09^{+ 0.24}_{-0.25}\,, \quad
\delta\varphi_4'= -0.41^{+ 1.29}_{-1.38}\,.
\ee
This yields relative contributions at the level of $\sim 20\%$ and $\sim 15\%$, respectively.
More specifically, the quantities $\varphi_i\delta\varphi_i$ \eqref{eq.def.delta} associated with the deformation parameters are
\be
\varphi_2\delta\varphi_2=-0.60^{+1.52}_{-1.58}\longrightarrow -0.45^{+1.52}_{-1.57}\qand
\varphi_4\delta\varphi_4= -17.93^{+55.92}_{-60.11}\longrightarrow -15.30^{+55.93}_{-60.00}  \,.
\ee
We find that, even under conservative assumptions, eccentricity induces shifts of approximately $25\%$ and $15\%$ relative to the central values of the deformation parameters. However, these shifts remain well within the current statistical uncertainties and do not represent an order-of-magnitude effect. In particular, eccentricity does not introduce corrections large enough to mimic or dominate the inferred deviations from GR. Therefore, its impact can be regarded as subleading in the present analysis.

%It can be used to estimate the magnitude of the ecc impact, and has little effect on the phase correction.
Although eccentric corrections do not follow the standard ppE frequency dependence, this comparison provides a useful order-of-magnitude estimate. We find that, within the observationally allowed eccentricity range, eccentricity-induced phase corrections are subdominant and are unlikely to mimic significant deviations from quasi-circular GR waveforms.

%Conclusion: It has little effect on deformation.
We therefore conclude that orbital eccentricity constitutes a subleading source of systematic uncertainty in the estimation of deformation parameters for GW170608, and does not qualitatively affect the conclusions of this work.

\section{Conclusions}
\label{sec:conclusions}

%{\color{red} our method and task }
The GW phase during the inspiral phase of binary BHs in models beyond GR is entirely determined by modifications to both the conservative and dissipative sectors of the system. We assume that the conservative sector is fully characterized by the BH solution within the modified model, and that the dissipative corrections enter at a higher PN order than the modified gravity corrections themselves. Under this assumption, we have calculated the modifications to the inspiral phasing induced by the deformation parameters of various BH spacetimes beyond GR.

%{\color{red} our result }

The constraints on the deformation parameters obtained from GW observations are in excellent agreement with the predictions of GR. The analysis of the deformation parameters for a range of alternative metrics reveals no significant deviations from GR within the 90\% confidence intervals. This consistency indicates that, at least for the current GW data, the standard Kerr solution remains a valid description of astrophysical BHs. The non-Kerr metrics considered in this study, which encompass various possible deformations in the spacetime geometry, are not required by the data to explain the observed GW signals. These results suggest that the current sensitivity of GW detectors is sufficient to rule out significant deviations from GR, but also highlight the potential for future observations to provide tighter constraints as detector sensitivity improves.

In addition to GW constraints, we also compared the results with those obtained from BH shadow imaging and X-ray reflection spectroscopy. The constraints from both these methods are consistent with our GW results, reinforcing the conclusion that deviations from the Kerr metric are not supported by current observational data. For most BH metrics, there are no studies reported in the literature to constrain these BH spacetimes with X-ray reflection spectroscopy, but when they are available, they are stronger than those from GWs found in this work. For the BHs in asymptotically safe gravity, GWs provide the constraint $\tilde{\omega} < 0.78$ while the constraint from X-ray reflection spectroscopy is $\tilde{\omega} < 0.05$ (both at 90\% confidence level). For the Kerr-Sen BH metric, the GW constraint is $r_\alpha < 0.58$ and the constraint from X-ray reflection spectroscopy is $r_\alpha < 0.011$. However, assuming that the deformation parameter has the same value for every BH, we could combine GW events together and get a stronger constraint on the deformation parameter from GWs~\cite{Cardenas-Avendano:2019zxd}.

We have also assessed the impact of orbital eccentricity as a potential source of systematic uncertainty. Using a leading-order, order-of-magnitude estimate, we find that, within the observationally allowed eccentricity range for GW170608, the associated phase corrections are subdominant compared to the inferred deviations. Consequently, eccentricity is unlikely to mimic or significantly bias the constraints on deformation parameters, and does not qualitatively affect the conclusions of this work.

In summary, the results presented in this study highlight the ability of GWs to constrain deviations from the Kerr metric. The consistency of our findings with BH shadow imaging and X-ray reflection spectroscopy further strengthens the case for GR as the dominant description of BH spacetimes. Looking ahead, upcoming advancements in GW detectors, as well as improvements in multi-messenger astronomy, will continue to refine these constraints and may allow us to explore even more exotic spacetime geometries.

\section*{Acknowledgments}
This work was supported by the National Natural Science Foundation of China (NSFC), Grant Nos.~W2531002, 12261131497, and 12250610185.

The manuscript is based upon work supported by NSF’s LIGO Laboratory, which is a major facility fully funded by the National Science Foundation (NSF), as well as the Science and Technology Facilities Council (STFC) of the United Kingdom, the Max-Planck-Society (MPS), and the State of Niedersachsen/Germany for support of the construction of Advanced LIGO and construction and operation of the GEO600 detector. Additional support for Advanced LIGO was provided by the Australian Research Council. Virgo is funded through the European Gravitational Observatory (EGO), by the French Centre National de Recherche Scientifique (CNRS), the Italian Istituto Nazionale di Fisica Nucleare (INFN) and the Dutch Nikhef, with contributions by institutions from Belgium, Germany, Greece, Hungary, Ireland, Japan, Monaco, Poland, Portugal, Spain. KAGRA is supported by the Ministry of Education, Culture, Sports, Science and Technology (MEXT), Japan Society for the Promotion of Science (JSPS) in Japan; National Research Foundation (NRF) and the Ministry of Science and ICT (MSIT) in Korea; Academia Sinica (AS) and National Science and Technology Council (NSTC) in Taiwan, China.

%\bibliographystyle{unsrt}
%\bibliography{cite.bib}

\end{document}